\documentclass[aps,prd,twocolumn,floatfix,superscriptaddress]{revtex4-2}
\usepackage{tabularx}
\usepackage{graphicx}
\usepackage{amsmath}
\usepackage{amssymb}
\usepackage{comment}
\usepackage{xcolor}
\usepackage{xspace}
\usepackage{multirow}
\usepackage{natbib}

\newcommand{\dfdplr}{\Delta f_{\rm dplr}}
\newcommand{\nruns}{N_{\rm runs}}
\newcommand{\niter}{N_{\rm iter}}
\newcommand{\fitalgoname}{\texttt{GBSIEVER}\xspace}
\newcommand{\tdia}{A\xspace}
\newcommand{\tdie}{E\xspace}
\newcommand{\tdit}{T\xspace}
\newcommand{\tdiamath}{{\rm A}}
\newcommand{\tdiemath}{{\rm E}}

\newcommand{\gbamp}{\mathcal{A}}
\newcommand{\intpar}{\kappa}
\newcommand{\fstat}{$\mathcal{F}$-statistic\xspace}

\DeclareMathOperator*{\argmin}{argmin}
\DeclareMathOperator*{\argmax}{argmax}

\begin{document}
\title{Resolving Galactic binaries in LISA data using particle swarm optimization and cross-validation}
\author{Xue-Hao Zhang}
\affiliation{Institute of Theoretical Physics \& Research Center of Gravitation, Lanzhou University, Lanzhou 730000, China}
\affiliation{Lanzhou Center for Theoretical Physics, Lanzhou University, Lanzhou 730000, China}
\affiliation{Morningside Center of Mathematics, Academy of Mathematics and System Science, Chinese Academy of Sciences, 55, Zhong Guan Cun Donglu, Beijing 100190, China}
\author{Soumya D.~Mohanty}
\affiliation{Dept.~of Physics and Astronomy, University of Texas Rio Grande Valley, One West University Blvd.,
Brownsville, Texas 78520, USA}
\affiliation{Morningside Center of Mathematics, Academy of Mathematics and System Science, Chinese Academy of Sciences, 55, Zhong Guan Cun Donglu, Beijing 100190, China}
\email{soumya.mohanty@utrgv.edu}
\author{Xiao-Bo Zou}
\affiliation{Institute of Theoretical Physics \& Research Center of Gravitation, Lanzhou University, Lanzhou 730000, China}
\affiliation{Lanzhou Center for Theoretical Physics, Lanzhou University, Lanzhou 730000, China}
\affiliation{Morningside Center of Mathematics, Academy of Mathematics and System Science, Chinese Academy of Sciences, 55, Zhong Guan Cun Donglu, Beijing 100190, China}
\author{Yu-Xiao Liu}
\affiliation{Institute of Theoretical Physics \& Research Center of Gravitation, Lanzhou University, Lanzhou 730000, China}
\affiliation{Lanzhou Center for Theoretical Physics, Lanzhou University, Lanzhou 730000, China}
%%%%%%%%%%%%%%%%%%%%%%%%%%%%%%%%%%%%%%%%%%%%%%%%%%%%%%%%%
\begin{abstract}
The space-based gravitational wave (GW) detector LISA is expected to
observe signals
from a large population of compact object binaries, comprised predominantly of
white dwarfs, in the Milky Way. Resolving individual 
sources from this population against its self-generated confusion noise poses a major data analysis problem. We present an iterative 
source estimation and subtraction method to address this problem based on the use of particle swarm optimization (PSO). In addition to PSO, a
novel feature of the method is the cross-validation of sources estimated from 
the same data using different signal parameter search ranges. 
This is found to
greatly reduce contamination by spurious sources 
and may prove to be a useful addition to any multi-source resolution method.
Applied to a recent mock data challenge, the method 
is able to find $O(10^4)$ Galactic binaries
across a signal frequency range of $[0.1,15]$~mHz, 
and, for frequency $\gtrsim 4$~mHz, reduces the residual data after subtracting out estimated signals to the instrumental noise level.

\end{abstract}
\maketitle

%%%%%%%%%%%%%%%%%%%%%%%%%%%%%%%%%%%%%%%%%%%%%%%%%%%%%%%%%%
\section{Introduction}
\label{sec:intro}
The gravitational wave (GW) window in the frequency range $[10,10^3]$~Hz has
now been opened using ground-based detectors: the LIGO~\cite{aasi2015advanced} and Virgo~\cite{acernese2014advanced} detectors have
observed several binary black hole mergers~\cite{LIGOScientific:2018mvr,Abbott:2020niy,Venumadhav:2019lyq} and one double neutron star merger~\cite{TheLIGOScientific:2017qsa}. 
Work is in progress to extend this success
to lower frequency ranges. The space-based Laser Interferometric Space Antenna (LISA) mission~\cite{Audley:2017drz}, to be launched around $2034$, will probe the $[10^{-4}, 1]$~Hz band. Key experimental technologies required for LISA have 
been demonstrated successfully in the LISA Pathfinder mission~\cite{Armano:2009zz}. LISA
may be joined by additional space-based missions currently under study, namely, 
Taiji~\cite{Guo:2018npi} and Tianqin~\cite{Luo:2015ght}. The exploration of the nanohertz frequency range, $[10^{-9}, 10^{-7}]$~Hz, is well underway
using pulsar timing arrays (PTAs)~\cite{hobbs2010international},
with  the
sensitivity of these searches expected to increase by orders of magnitude~\cite{ Wang:2016tfy}
as next-generation radio telescopes~\cite{Hobbs:2014tqa,Smits:2008cf} come 
online over the coming decade.

The planned configuration of LISA~\cite{Audley:2017drz} is a set of three  satellites in heliocentric orbits, each one housing freely falling proof masses, that will maintain a triangular formation
trailing the Earth 
by about $20^\circ$. At the lowest order of approximation, the 
triangle is rigid and equilateral with an arm length of $2.5\times 10^6$~km. It is tilted with respect to the ecliptic and rotates around its centroid with a period of $1$~year. 
GW induced fluctuation in the arm lengths will 
be measured using the technique of 
time delay interferometry (TDI)~\cite{Tinto:2014lxa} that
linearly combines time-delayed readouts of frequency shifts in the light exchanged
by the satellites. There are 
several such TDI combinations, of which the so-called \tdia, \tdie, \tdit 
combinations have mutually uncorrelated instrumental noise.

Since LISA is a nearly omnidirectional detector, GW signals from sources distributed
across the sky will appear simultaneously and additively in TDI data. Predominant 
among these are expected to be long-lived signals from a variety of 
compact object binaries, including
(i) several 
Massive Black Hole Binaries (MBHBs)~\cite{Katz:2019qlu} with components masses in the $[10^3, 10^6]$~$M_\odot$ range, (ii) a few extreme mass ratio systems~\cite{Babak:2017tow} of stellar mass black holes in orbit around massive ones, and (iii) $O(10^8)$ Galactic binaries (GBs) within the Milky Way comprised primarily of white dwarfs~\cite{Nelemans:2001hp}. 

Disentangling a myriad of signals from multiple sources of different types,
i.e., multisource resolution,
is a major data analysis problem for LISA. 
The LISA community has organized a series of challenges to encourage the development of data analysis methods for addressing this humongous task.
The first to the fourth of these 
were called Mock LISA Data Challenges (MLDCs)~\cite{Arnaud_2007,Babak:2007zd,babak2010mock} 
and the subsequent ones are part of a series called
 LISA Data Challenge (LDC)~\cite{ldcpage}. 
The first LDC (LDC1)
consists of several sub-challenges, of which the fourth (LDC1-4) pertains to GB 
resolution. 

While the signal from a GB in its
source frame is quite simple in form -- a nearly constant amplitude linear chirp with an 
increasing or decreasing 
secular frequency drift depending on mass transfer, or lack thereof, between the components -- its spectrum is broadened considerably 
due to the frequency and amplitude modulations imposed by 
the motion of LISA. The resulting increase in the overlap of the signals,
accompanied by their sheer number, makes the GB resolution problem
especially challenging. In particular, GBs should be
so numerous at low frequencies ($\lesssim 3$~mHz) that the confusion noise from the blending of their signals is expected to dominate the instrumental noise. Thus, not only do individual sources need to be differentiated from each other but they also need to be resolved against
 the background of confusion noise, with the latter defined by, as well as 
 affecting, the data  analysis method used.

Quantifying the performance of multisource 
resolution methods is not as straightforward as the case of a single source search. 
The sources found by any method, called {\em reported sources} in this paper, will
generally consist of some that correspond to true sources while others that do not.
Differentiating between them 
is non-trivial since no reported source will have an exact match, in either 
its signal waveform or source parameters, with a true source. 
Past data challenges have employed a test, described in detail later, for this purpose that is based on the cross-correlation of reported and true signal waveforms. 
In the following, we refer to reported sources that pass such a 
test as {\em confirmed sources}
and their number relative to that of reported sources
as the {\em detection rate}. Note that in some methods,
the reported sources themselves may be 
filtered out from 
a much larger set of {\em identified sources} by applying some cuts to
reject {\em spurious sources}. 

A variety of data analysis methods have been proposed for the GB resolution problem. Some  (e.g.,~\cite{Mohanty:2005ca}) remain proof of principle studies 
while others were quite mature even by the first data challenge. The methods that have emerged at the top of the pack through the data challenges are as follows. (i)BAM~\cite{Crowder_2007}
combined simulated annealing and Markov Chain Monte Carlo (MCMC), in a block-wise frequency search to report $\simeq 20,000$ sources in MLDC-2.1 data that had $26$~million GBs. From Fig.1 in~\cite{Babak:2007zd}, we estimate  that it achieved a detection rate of $\approx 85\%$.  (ii) An extension of BAM~\cite{Littenberg:2011zg} incorporating trans-dimensional MCMC was tested on the training data from MLDC-4 and reported $\simeq 9,000$ sources in the $[0,10]$~mHz band with a  detection rate of $\simeq 90\%$. (The loss in performance relative to BAM was attributed to  an additional signal parameter, namely, the secular frequency drift, that was missing in MLDC-2.1.) (iii) A deterministic search with iterative source subtraction proposed in~\cite{Blaut:2009si} reported $\simeq 12,000$ sources in MLDC-3.1 data containing $60$~million GBs. (The viability of the deterministic and iterative source subtraction approach for GB resolution was demonstrated earlier  in~\cite{cornish2003lisa}.) A new front in GB resolution has been opened recently~\cite{Littenberg_time_evolving_GB} where the goal is to create a time-evolving source catalog as the data from LISA accumulates. In contrast, the  methods listed  above use the entire data from a fixed observation period.
 
 In this paper, we introduce a new iterative source subtraction 
 method for the GB resolution problem
 and demonstrate it on LDC1-4 data as well as
 data, that we call MLDC-3.1mod, obtained by adding the GBs used in MLDC-3.1
 to the noise realizations in LDC1-4. By taking out possible differences in noise characteristics between
 the two challenges from the equation, the latter allows a
 more equitable assessment of the effects of the much larger number of sources
 in MLDC-3.1. Among the key novel features of our method are the use of particle swarm optimization (PSO)~\cite{kennedy1995particle} for accelerating the baseline task of single source estimation, and the rejection of spurious sources by cross-validating identified sources found 
 with different search ranges for the secular
 frequency drift parameter. We call the method \fitalgoname: Galactic Binary Separation by Iterative Extraction and Validation using Extended Range. 
 
 Depending on the cuts that filter out reported 
 sources from identified ones, the number of reported sources for LDC1-4 data from 
 \fitalgoname falls between $9,291$ and $12,270$ in
 the $[0.1,15]$~mHz band 
 with an overall detection rate between $91.19\%$ 
 and $84.28\%$, respectively.
 For MLDC-3.1mod the numbers are $9,387$ and $12,044$
 with detection rates of $90.26\%$ and $84.16\%$, respectively.
 These results are
 comparable to the ones for the methods listed earlier.
The numbers above are just snapshots of an
extensive performance analysis of \fitalgoname contained in this paper 
that also includes an assessment of parameter estimation errors. 

The rest of the paper is organized as follows. 
The challenge data used in this paper are described in Sec.~\ref{sec:data}. 
The baseline
single-source detection and parameter estimation method is described in Sec.~\ref{sec:snglsrcstat}. \fitalgoname is described in 
Sec.~\ref{sec:method}, 
followed by the
results in Sec~\ref{sec:results}. 
Our conclusions and prospects for future work  are 
presented in Sec.~\ref{sec:conclusions}.

%%%%%%%%%%%%%%%%%%%%%%%%%%%%%%%%%%%%%%%%%%%%%%%%%%%%%%%%%%%%%%
\section{Challenge data description}
\label{sec:data}
We begin with a brief outline of LDC1-4 data, limited to establishing the notation used 
in the rest of the paper. 
For a plane GW, with polarizations $h_+(t)$ and $h_\times (t)$ (in the TT-gauge), emitted by
a source located in the Solar System Barycentric (SSB) frame at
ecliptic latitude, $\beta$, and ecliptic longitude, $\lambda$, 
the response of a single arm indexed by $l$ and having length $L$ is,
\begin{align}
y_{slr}(t) & = \frac{\Phi_l\left(t^\prime - \widehat{k}\cdot  \overline{R}_s(t^\prime)\right)-
\Phi_l\left(t-\widehat{k}\cdot \overline{ R}_r(t)\right)}{2(1-\widehat{k}\cdot \widehat{n}(t))}\;,\\
\Phi_l(t) & = \sum_{a=+,\times} F_{a,l}(t;\lambda,\beta,\psi)h_a(t)\;,
\end{align}
where, $t^\prime = t-L/c$, $\overline{R}_{s}$ and
$\overline{R}_r$ are, respectively, the SSB frame positions of the satellites (indexed by $s$ and $r$) sending and receiving light along this arm, $\widehat{n}(t)$ is the unit vector along $\overline{R}_r(t) - \overline{R}_{s}(t^\prime)$,  $\widehat{k}$ is the unit vector along the GW propagation direction,  $F_{+,l}$ and $F_{\times,l}$ are the antenna pattern functions for the arm, and  $\psi$ -- the polarization angle -- defines the orientation of the wave frame axes in the plane
orthogonal to $\widehat{k}$.
The time dependence of the antenna patterns is induced by the 
orbital motion of LISA. The GW signal in each TDI combination is obtained by combining the single
arm responses with  prescribed time delays. (Further details about TDI data 
generation are in the LDC1 manual~\cite{ldcpage}.)

For a GB, $h_{+,\times}(t)$ are linear chirps, 
%\dzcomment{This is the same as which in LDC manual. I've changed $\omega(t)$ to $\Phi(t)$.}
\begin{align}
h_+(t) & = \gbamp \left(1 + \cos^2\iota\right) \cos \Phi(t)\;,\\
h_\times(t) & = -2 \gbamp \cos\iota \sin \Phi(t)\;,\\
\Phi(t) & = \phi_0 + 2\pi f t + \pi \dot{f} t^2\;,
\end{align}
that are parametrized by the overall amplitude $\gbamp$, initial phase $\phi_0$ at the start of observation, inclination $\iota$ of the GB orbit to the line of sight from the 
SSB origin,
carrier frequency $f$, and secular frequency drift $\dot{f}$. (For a GB, $\psi$ specifies the orientation of the orbit's projection on the sky.)
The majority of GBs in LDC1-4 data
have frequencies in the $\approx [0.1,15]$~mHz range, with  only 18 having $f > 15$~mHz,
and the
source number density in frequency increases as one moves to lower frequencies.  

\begin{figure}[h]
\centering
\includegraphics[width=\columnwidth]{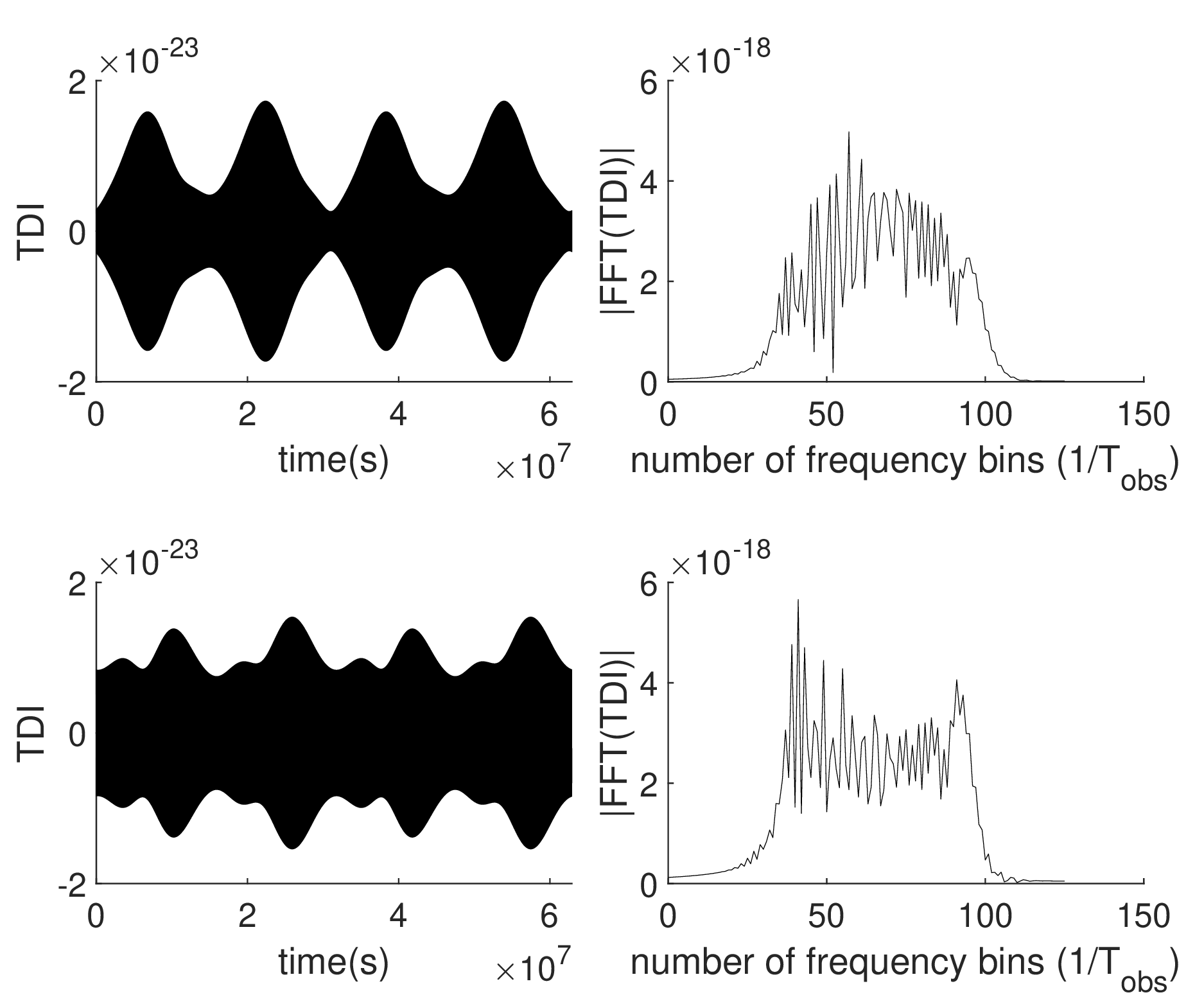}
\caption{TDI A (top row) and E (bottom row) signals corresponding to a single GB with $f\simeq 5$~mHz and $T_{\rm obs}=2$~yr in (left column) the time and (right column) the Fourier domain. 
The 
Fourier frequency is expressed in units of $1/T_{\rm obs}$.}
\label{fig:GBsig}
\end{figure}

 Fig.~\ref{fig:GBsig} illustrates
 \tdia and \tdie signals, in both the time and Fourier domains, from a single GB. (The T combination is 
ignored from here on because the GW signal 
in it is highly attenuated at low frequencies.) The Doppler shift arising from
the orbital motion of LISA and 
 the time dependence of its antenna patterns induce periodic modulations in
 the instantaneous frequency and amplitude of an observed GB signal.
 Over an observation period $T_{\rm obs}=2$~years,
the energy of a GB 
signal with frequency $f$ (Hz) is spread by Doppler modulation over a frequency 
range $\dfdplr$ that is a factor of 
$\approx 10^4\times f$ 
larger than $1/T_{\rm obs}$, 
the minimum spacing of Fourier frequencies. This not only reduces the peak 
Fourier domain amplitude
of GB signals substantially but also increases their overlap, thereby making the 
problem of resolving them very challenging. 

Each TDI combination in LDC1-4 contains the sum of  signals from a set
of $30$~million GBs drawn randomly from an astrophysically realistic 
population model~\cite{Nelemans:2001hp}. (The corresponding number for MLDC-3.1 is $60$~million.)
To this collective signal is added a
realization of a pseudo-random sequence that models the 
instrumental noise. 
The software, called \texttt{LISACode}~\cite{Petiteau:2008zz}, that generates 
 LDC TDI data
also provides the respective noise realizations. Both LDC1-4 and MLDC-3.1 
disclose the 
true parameters of the GBs used in synthesizing  
their respective challenge data.  We generated the TDI combinations corresponding to
the GB parameters in MLDC-3.1 and added them to the \texttt{LISACode}
 noise realizations. As mentioned in Sec.~\ref{sec:intro}, we
call the resulting data MLDC-3.1mod. 

\begin{figure}
\centering
\includegraphics[width=\columnwidth]{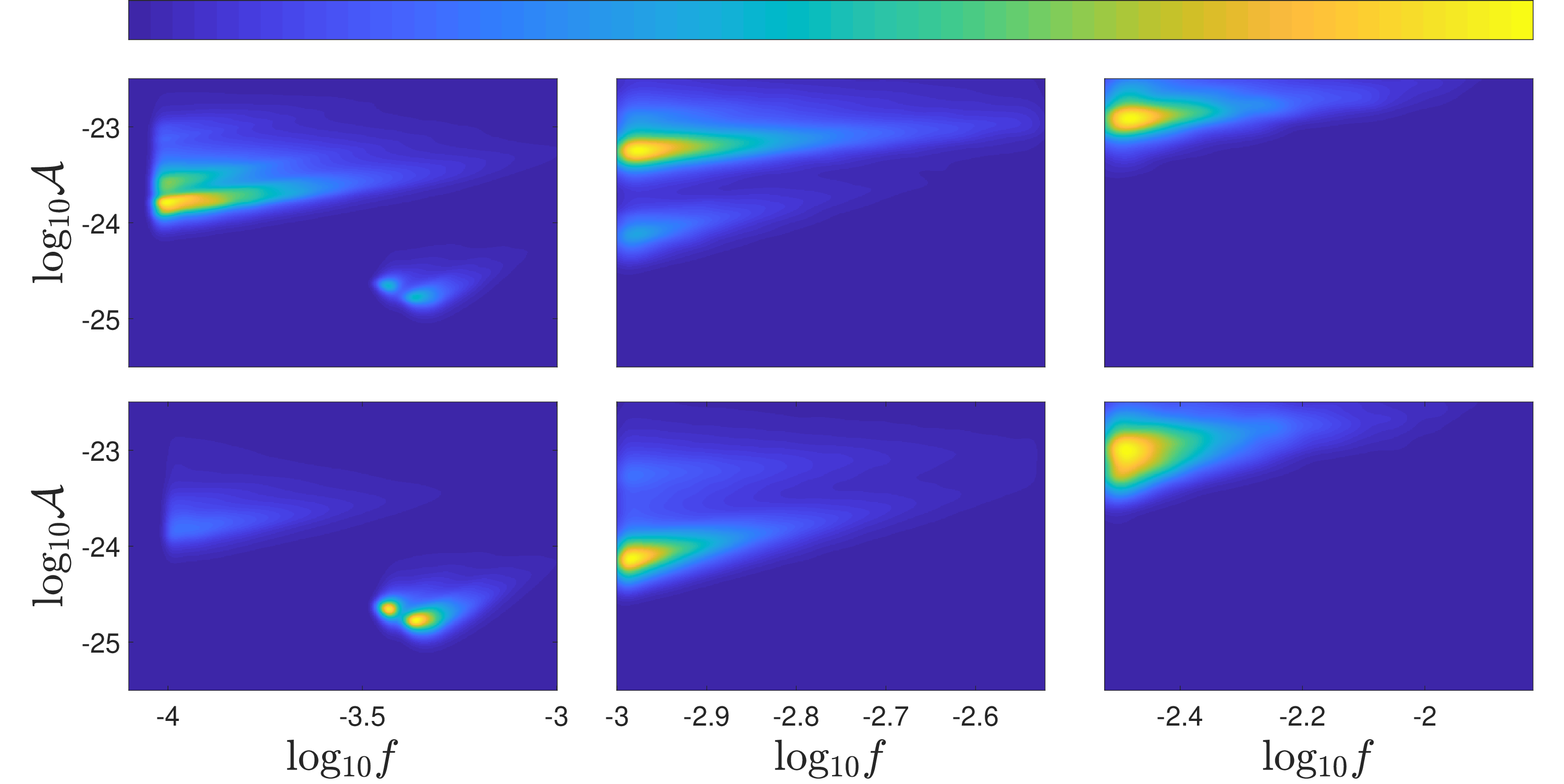}
\caption{Kernel density estimate of the joint distribution of amplitude, $\mathcal{A}$,
and frequency, $f$ (Hz), for LDC1-4 (top row) and MLDC-3.1 (bottom row) sources. The range of values on the Y-axis is the same in each row and shown only for the leftmost column for clarity. The estimated density in each panel is color coded 
according to the map shown at the top, with values increasing
from left to right. The color coding of density values is the same for LDC1-4 and MLDC-3.1 within each column of panels (i.e., frequency interval) but differs across the columns. This allows the extreme values of density that occur in the combined frequency range to be displayed simultaneously.}
\label{fig:A-f}
\end{figure}

Fig.~\ref{fig:A-f} compares the distributions of the amplitude parameter in LDC1-4 and MLDC-3.1. Even
though the latter has twice the number of sources, the 
majority of them are low amplitude ones. Hence, one does not 
expect to resolve a significantly larger number of GBs in MLDC-3.1 (or MLDC-3.1mod).
 The range of true $\dot{f}$ in these datasets is
relevant for the cross-validation step in \fitalgoname: In LDC1-4, $-3.3\times 10^{-17}\lesssim \dot{f} \lesssim 6.8\times 10^{-16}$~${\rm Hz}^2$ for $f<4$~mHz and $-3.0\times10^{-15}\lesssim \dot{f}\lesssim 3.9\times 10^{-14}$ ~${\rm Hz}^2$ for $f\in [4,15]$~mHz; In MLDC-3.1, $-3.8\times 10^{-17}\lesssim \dot{f} \lesssim 1.1\times 10^{-15}$~${\rm Hz}^2$ for $f<4$~mHz and $-2.3\times10^{-14}\lesssim \dot{f}\lesssim 7.7\times 10^{-14}$ ~${\rm Hz}^2$ for $f\in [4,15]$~mHz.
%%%%%%%%%%%%%%%%%%%%%%%%%%%%%%%%%%%%%%%%%%%%%%%%%%%%%%%%%%%%%%
\section{F-statistic and its implementation}
\label{sec:snglsrcstat}

Each TDI time series in LDC1-4 is uniformly sampled with a sampling interval  $\Delta = 15$~sec.  With $\overline{x}$ denoting a row vector, the data from a TDI combination $I\in \{\tdiamath,\tdiemath\}$ is given by 
\begin{eqnarray}
\overline{y}^I & = & \overline{S}^I + \overline{n}^I\;, 
\end{eqnarray}
where $\overline{S}^I$ denotes the collective GW signal from all the GBs,
any one of which will be denoted by $\overline{s}^I$,
and $\overline{n}^I$ denotes the instrumental noise realization. 
For LDC1-4,
$\overline{y}^I\in \mathbb{R}^N$ with $N = 4194304$, corresponding
to $T_{\rm obs}\approx 2$~yr. 
The instrumental noise is modeled by a Gaussian, stationary, stochastic process with 
power spectral density (PSD) $S_n^I(\nu)$ at Fourier frequency $\nu$.  

%%%%%%%%%%%%%%%%%%%%%%%%%%%%%%%%%%%%%%%%%%%%%%%

\subsection{Parameter estimation: single source}
\label{sec:fstat}
\fitalgoname uses iterative estimation of signals in which each step estimates the parameters of a GB signal using
 maximum likelihood estimation (MLE) under the assumption
 that the data contains only a single source. The estimated signal is subtracted
from the data and the process is repeated on the residual.
(The formalism for MLE below 
closely follows the one in~\cite{Blaut:2009si} to which we refer
the reader  for more details.)

The GW signal from a single GB can be expressed, schematically, as 
\begin{align}
    \overline{s}^I(\theta) & = \overline{a}\,{\bf X}^I(\intpar)\;,
\end{align}
 where $\overline{a}=(a_1, a_2, a_3, a_4)\in \mathbb{R}^4$ and ${\bf X}^I(\intpar)$ is a $4$-by-$N$ matrix of {\em template} waveforms. 
The former, called the set of {\em extrinsic} parameters, is a reparametrization of the four parameters $\gbamp$, $\phi_0$, $\psi$, and $\iota$. The latter depends on the remaining set, $\intpar = \{\lambda, \beta, f, \dot{f}\}$,
of the so-called {\em intrinsic} parameters.
The complete set of 
parameters is denoted by $\theta = \{\overline{a},\intpar\}$. 
We make a parenthetical remark here that \fitalgoname uses the 
expressions in~\cite{Blaut:2009si} for generating template waveforms while
the signals in LDC1-4 are generated by the \texttt{FastGB} code~\cite{Cornish:2007fb}. However, we have verified that the two are in good agreement for the \tdia and
\tdie combinations in the frequency range of interest here.

 The point estimate of the parameters $\theta$ in an iteration 
 is obtained as,
\begin{align}
    \widehat{\theta} & =  \argmin_{\theta} \sum_{I\in\{A,E\}}\left(\| \overline{y}^I - \overline{s}^I(\theta)\|^I\right)^2\;,
    \label{eq:mletheta}
\end{align}
where 
it is understood that $\overline{y}^I$ is the residual from the 
previous step. Here,
$\|.\|^I$ is the norm induced by the inner product,
\begin{align}
    \langle \overline{x},\overline{z}\rangle^I & = \frac{1}{Nf_s}(\widetilde{x}./\overline{S}_n^I)\widetilde{z}^\dagger\;,
    \label{eq:innprod}
\end{align}
where
$f_s = 1/\Delta$ is
the sampling frequency, $\widetilde{x}^T  = {\bf F}\overline{x}^T$ is the discrete Fourier transform (DFT) of $\overline{x}$ defined by $F_{kl} = \exp(-2\pi i k l/N)$,  
$./$ denotes element-wise division, and $\overline{S}_n$ contains the
samples of $S_n(\nu)$ at the DFT frequencies. 

Let $\overline{A}(i)$ denote  row $i$ and $A(i,j)$ 
 the element in  row $i$ and 
 column $j$
of a matrix ${\bf A}$. Let
${\bf U}^I$ denote the column vector with 
\begin{align}
    U^I(i) &= \langle \overline{y}^I, \overline{X}^I(i)\rangle^I\;,
    \label{eq:umat}
\end{align}
and ${\bf W}$
denote the matrix with
\begin{align}
    W^I(i,j) &= \langle \overline{X}^I(i),
                            \overline{X}^I(j)\rangle^I\;.
    \label{eq:wmat}
\end{align}
Then, the minimization problem in Eq.~\ref{eq:mletheta} can be recast
as a maximization,
\begin{align}
    \widehat{\theta} & =  \argmax_\theta \sum_{I\in\{A,E\}}\left(
            \overline{a}{\bf U}^I - \frac{1}{2}\overline{a}{\bf W}^I\overline{a}^T
    \right)\nonumber\\
    & = \argmax_\theta \left( \overline{a}{\bf U} - \frac{1}{2}\overline{a}{\bf W}\overline{a}^T\right)\;,
\end{align}
where ${\bf W} = \sum_I {\bf W}^I$ and ${\bf U} = \sum_I {\bf U}^I$.
For fixed $\kappa$, the maximization over $\overline{a}$ is trivial,
\begin{align}
    \widehat{a}^T & = {\bf W}^{-1}{\bf U}\;.
    \label{eq:amp_est}
\end{align}
The MLE estimate of $\kappa$ is then given by 
\begin{align}
    \widehat{\kappa} & =  \argmax_\kappa \mathcal{F}(\kappa)\;,
\end{align}
where
\begin{align}
       \mathcal{F}(\kappa) & = {\bf U}^T {\bf W}^{-1}{\bf U}\;
\end{align}
is widely known in the GW literature as the \fstat.

%%%%%%%%%%%%%%%%%%%%%%%%%%%%%%%%%%%%%%%%%%%%%
\subsection{White noise approximation}
\label{sec:bandlimit}
In common with other GB resolution methods, 
iterative estimation and subtraction of sources 
in \fitalgoname is carried out  
in restricted frequency bands.
Let  $[\nu_i, \nu_i+B_i]$ denote the frequency
limits demarcating band $i$.
If $B_i$ is kept sufficiently small, the noise
PSD $S_n(\nu)$ within the band can be treated as  
approximately constant, $S_n(\nu)\approx S_n^{(0)}(\nu_i)$. 
In \fitalgoname,
$S_n(\nu)$ is estimated
from the instrumental noise realization produced by \texttt{LISACode}~\cite{Note1} and
$S_n^{(0)}(\nu_i)$ is set to its mean value in $[\nu_i, \nu_i+B_i]$.  Under this approximation, the inner product defined in 
Eq.~\ref{eq:innprod} reduces to the Euclidean one,
\begin{align}
    \langle \overline{x},\overline{z}\rangle & = \frac{1}{S_n^{(0)}(\nu_i)f_s}\overline{x}\,\overline{z}^T\;.
    \label{eq:wgninnprod}
\end{align}
(Here, we have dropped the TDI index for clarity.) When computing
the above inner product, both $\overline{x}$ and $\overline{z}$ 
should 
be bandlimited to $[\nu_i, \nu_i+B_i]$ in keeping with the approximation of constant
in-band PSD. For a GB template, this condition holds implicitly 
as long as 
$B_i \gg \dfdplr$
and the frequency $f$ of the template 
is sufficiently far away from the band edges. For the data $\overline{y}$, this condition has to be enforced explicitly by using either a time domain bandpass filter or by windowing in the Fourier domain.  
%%%%%%%%%%%%%%%%%%%%%%%%%%%%%%%%%%%%%%%%%%%%

\subsection{Signal to noise ratio and correlation}
\label{sec:snr}
In the case where only a single GB source is present in the data,
the performance of \fstat is 
governed 
 by the optimal signal to noise ratio (SNR), defined as
\begin{align}
    {\rm SNR}^2 & = \sum_{I}\left(\|\overline{s}^I(\theta)\|^I\right)^2\;.
    \label{eq:snr}
\end{align}
On the other hand, the performance of
\fstat in the case of 
multiple GB signals depends not only on their SNR values
but also on the degree to which they interfere with each other
in the parameter estimation process. One measure of this is 
the {\em correlation coefficient} between pairs of signals,
\begin{align}
    R(\theta,\theta^\prime) & = \frac{C(\theta,\theta^\prime)}{\left[C(\theta,\theta)C(\theta^\prime,\theta^\prime)\right]^{1/2}}\;,\\
    C(\theta,\theta^\prime) & = \sum_{I}\langle \overline{s}^I(\theta),\overline{s}^I(\theta^\prime)\rangle^I\;.
    % \frac{\sum_{I}\langle \overline{s}^I(\theta),\overline{s}^I(\theta^\prime)\rangle^I}{\sum_{I}\left(\|\overline{s}^I(\theta)\|^I\right)^2 \| \overline{s}^I(\theta^\prime)\|^I }\;. 
\end{align}
 One expects that with an increase in the fraction of signals with high mutual 
 correlations in a given set, the confusion noise from
 their blending will become stronger.

The correlation coefficient is also  used to 
match a reported source to a true one when analyzing simulated data (c.f., Sec.~\ref{sec:metrics}). In this 
context, it should be noted that $R(\theta,\theta^\prime)$ 
only measures similarity in the shapes of signal waveforms and
ignores differences in their SNR. Hence, to test for a match, the 
correlation coefficient needs to be supplemented with some SNR-based criterion.
%%%%%%%%%%%%%%%%%%%%%%%%%%%%%%%%%%%%%%%%%%%%

\subsection{Particle swarm optimization}
\label{sec:pso}
The maximization of \fstat over the space of intrinsic parameters $\kappa$  is a challenging high-dimensional, non-linear, and 
non-convex optimization problem. \fitalgoname accomplishes this task using PSO, a 
nature-inspired stochastic optimization method modeled
after the  behavior of a cooperative of freely moving 
agents, such as a flock of birds, that can efficiently
search an area for the best value
of a distributed 
quantity, such as food. 

The basic idea in PSO is to 
mimic this behavior by evaluating  a given fitness function -- \fstat in our
case -- at multiple locations, the locations being particles
and the entire set of locations being the swarm. The particles  explore the search space randomly for better fitness values following iterative rules
that incorporate
cognizance of the behavior of the swarm as a whole. If a particle 
chances upon a good fitness value, the swarm eventually converges to its location and refines the fitness value further. In the
process of converging, however, the swarm can find a better
fitness value than the current one and shift its attention elsewhere, allowing it to escape local optima.

At present, PSO  more properly refers to a 
family of stochastic optimization algorithms, i.e., a metaheuristic~\cite{engelbrecht2005fundamentals}, with considerable variations between them
but organized around the basic idea outlined above. In the particular variant used in \fitalgoname, 
called {\em local best} (lbest) PSO~\cite{eberhart1995new}, the propagation 
of information within the swarm is deliberately slowed down, by splitting the swarm into 
overlapping cliques, to
extend the exploration phase.
A description of lbest PSO 
and, more broadly, a pedagogical introduction 
to using PSO in statistical regression problems can be found in~\cite{Mohanty2019}. 

Convergence to the global optimum is not guaranteed for 
 practical stochastic optimization methods, including PSO, 
 and extracting good performance on any given problem almost
always requires tuning the parameters of such a method. 
Remarkably, however, the  settings for the core parameters of
lbest PSO are observed to be fairly robust across
a wide range of problems~\cite{4223164}.
In fact, the PSO parameters in \fitalgoname are kept
 the same 
as in~\cite{normandin2018particle}, where
the optimization problem is the very different one of binary inspiral search in
ground-based GW detectors. Typically, tuning is required only for the number of iterations,  $\niter$,  until
termination of the search, and a hyper-parameter, $\nruns$, that specifies the number of independent
parallel runs of PSO on the same fitness function. 
The final  value of \fstat and its location is taken from
the run with the the best terminal fitness value. 

The tuning of $\niter$ and $\nruns$ in \fitalgoname was 
performed empirically
using simulated data realizations containing one to a few GB signals
spanning a wide range of SNRs. These parameters are deemed 
well-tuned~\cite{normandin2018particle} if 
the \fstat value found by PSO is higher than the one for the true 
signal parameters in a sufficiently large fraction of data realizations.
Following this procedure, the settings
$\niter = 2000$ and $\nruns = 6$ were
found to give good performance across all
frequency bands for 
sources with ${\rm SNR}\gtrsim 7$.

%%%%%%%%%%%%%%%%%%%%%%%%%%%%%%%%%%%%%%%%%%%%%%%%%%%%%%%%%%%%%%
\section{Description of GBSIEVER}
\label{sec:method}
Sec.~\ref{sec:snglsrcstat} described how the search for a single source in a single frequency band is carried out in \fitalgoname.
In this section we describe the key implementation details involved in the complete analysis
across multiple bands.
(For reference in the following, see Sec.~\ref{sec:intro} for
the definitions of the sets of sources
termed 
identified, reported, and confirmed, where each is a subset of the preceding.)
%%%%%%%%%%%%%%%%%%%%%%%%%%%%%

\subsection{Handling edge effects}
\label{sec:edge}
When using bandlimited data for source identification, high SNR  sources that 
happen by chance to fall near a band edge can generate a cluster of spurious estimated 
sources. This is caused by the
leakage of spectral power across the edge creating
 a false source that, when subtracted, injects a new spurious source into the data. This sequence of misidentifications perpetuates itself as the spectral power of the spurious sources leaks back and forth across the edge.
 The mitigation of such edge effects in \fitalgoname closely follows the prescription in~\cite{Crowder:2006eu,Littenberg:2011zg}: (a) Data is bandlimited in the Fourier domain using a  window function that tapers off at the band edges.   (b) Within a band, sources found 
 in the vicinity of its edges are rejected. (c) Adjacent bands are overlapped. (In~\cite{Crowder:2006eu}, the window 
 was rectangular but 
the noise PSD in each band was modified to have higher values near the edges.) 

\begin{figure}
\centering
\includegraphics[width=\columnwidth]{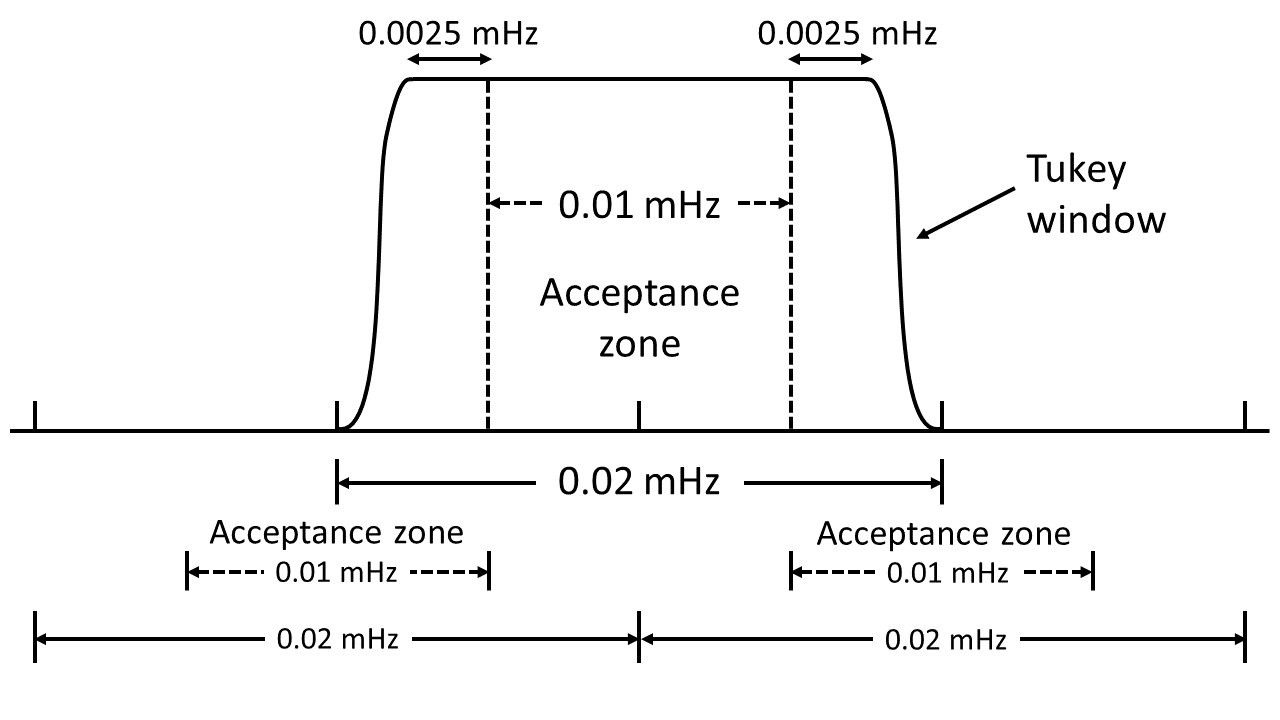}
\caption{ Schematic diagram of the windowing scheme. In this example, a  
 Tukey window bandlimits the data
in the Fourier domain to $0.02$ mHz. Only the 
estimated sources that belong to the central $0.01$~mHz interval, called
the acceptance zone, are retained as identified sources. The acceptance zone is narrower than the flat part of the Tukey window that covers $75\%$ of the full band. Adjacent 
$0.02$~mHz bands, having identical Tukey windows, are overlapped such that their 
acceptance zones are contiguous.}
\label{fig:accptWin}
\end{figure}

Fig.~\ref{fig:accptWin} illustrates the specifics of our approach.
Data is bandlimited in the Fourier domain using a Tukey window. 
At the end of  
the iterative subtraction process in a given band,
only the estimated sources  within an {\em acceptance zone} survive as identified sources while the rest are rejected. 
Adjacent bands are overlapped such that their
 acceptance zones are contiguous, allowing genuine
sources rejected in one band to be recovered in the acceptance zone of 
another. 

\begin{figure}
\centering
\includegraphics[width=\columnwidth]{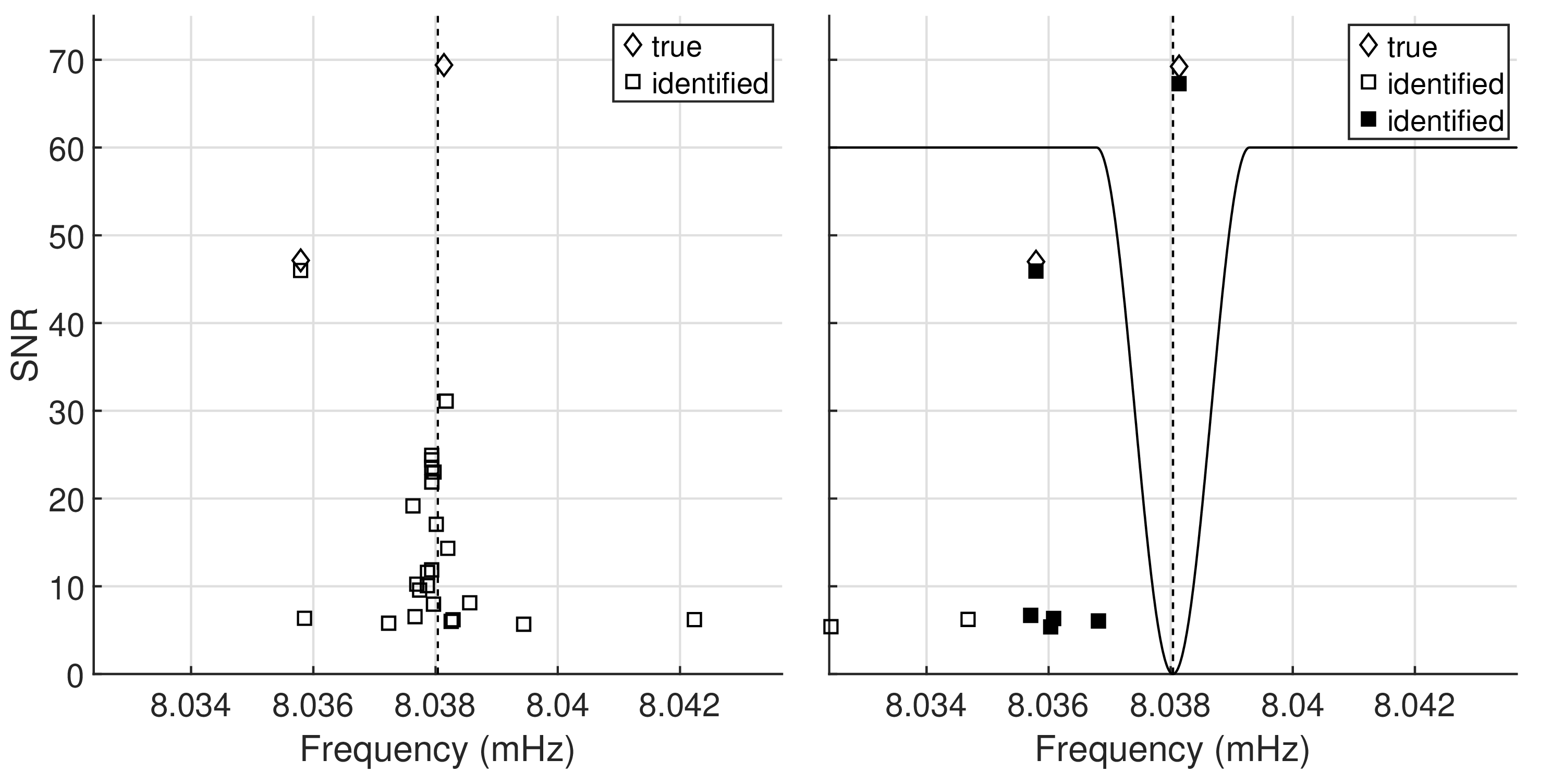}
\caption{ The edge effect and its mitigation with overlapped windowing. Each panel shows the halves of two adjacent non-overlapping 
bands, each $\approx 0.01$~mHz wide,  along with their common edge (dashed line). The SNR and frequency of true (diamonds) and identified (squares) sources  are shown for the case of (left panel)  a
 rectangular window with an acceptance zone spanning the whole band, and (right panel) 
a Tukey window (solid line) with an acceptance zone that is narrower than its flat part.
True sources that fall outside the acceptance zones of the bands shown here
are recovered (filled squares) in the acceptance zone of the band (not shown) centered
on the common edge. }
\label{fig:EdgeEffect}
\end{figure}

Fig.~\ref{fig:EdgeEffect} illustrates the edge effect and its
mitigation by overlapped Tukey windows.
For a rectangular window and an acceptance zone 
spanning the whole band, the presence of a true source
close to a band edge triggers a cluster of spurious sources
as described earlier. 
The tapering of the Tukey window suppresses this effect  substantially, thereby reducing the burden 
of dealing with spurious sources in subsequent stages of \fitalgoname. 

%%%%%%%%%%%%%%%%%%%%%%%%%%%%%

\subsection{Undersampling and data length reduction}
\label{sec:undersampling}
Computational efficiency in the calculation of the \fstat is of utmost importance in \fitalgoname 
since it must be evaluated a large number of times.
This is achieved by evaluating all 
inner products (c.f., Eq.~\ref{eq:wgninnprod}) using the method 
of {\em undersampling}~\cite{kester2003mixed} outlined below. 

For a given band $[\nu,\nu + B]$, the DFT, $\widetilde{y}$, of the data is windowed as described in Sec.~\ref{sec:edge}. The
inverse DFT of the windowed data is then 
 downsampled below the minimum Nyquist rate, $2(\nu+B)$, by retaining  a regularly spaced 
 subset of the samples. This direct sub-Nyquist rate sampling leads to
 aliasing error~\cite{bracewell1986fourier}, but if the new sampling frequency
 is $f_s/d$, where $d$ 
 is a positive integer such 
 that $2(\nu+B)/n\leq f_s/d\leq 2\nu/(n-1)$ for an integer $n\in [1,(\nu+B)/B]$, its effect is to create 
 a copy of the bandlimited spectrum in the 
 baseband $[0,(f_s/d)/2]$. Thus, this clever use of aliasing error
 preserves the information in the original
 data while reducing the number of samples.
 Choosing the largest $n$ compatible with the 
 bounds on $f_s/d$  yields the maximum allowed undersampling factor, $d$. 
As an example, for $\nu = 5$~mHz, $B = 0.02$~mHz, and $f_s = 1/15$~Hz (the default sampling frequency of LDC1-4 data), undersampling
reduces the number of samples by a factor $d=1414$. 
In contrast, if the data is simply downsampled to the minimum Nyquist rate, the reduction achieved is only
a factor of $6$. The above procedure is followed for each band
to generate the corresponding undersampled
data vector that is then stored for subsequent processing.

The principal benefit of undersampling is the
reduced computational cost of generating  templates (c.f., Sec.~\ref{sec:fstat}) on-the-fly in PSO since they need only be computed at the same time instants 
as the 
undersampled data. Since, unlike the data, templates are not explicitly bandlimited, the direct undersampling of  
a template incurs an error but it is negligible unless
the frequency parameter, $f$, 
lies within $\dfdplr$ of a band edge. However,  with a frequency this close to a band edge, the associated estimated source  
would be discarded anyway in the overlapped windowing scheme.
Hence, for all practical purposes,
the error in template waveforms due to undersampling  is ignorable.
%%%%%%%%%%%%%%%%%%%%%%%%%%%%%

\subsection{Termination rule}
\label{sec:stoprules}
The rule for terminating iterative source subtraction in a band can have a 
significant effect on the number and nature of identified sources. 
In \fitalgoname, termination happens when (i) a specified number, $N_{\rm term}$, of sources have been identified, or (ii) all the estimated source SNRs in $5$ consecutive 
iterations fall below a specified threshold $\eta_{\rm end}$. 
The second criterion accounts for the fact that identified sources are not 
found in a strictly descending order of 
SNR, especially as one approaches lower values, making termination based on 
 a single instance of ${\rm SNR}< \eta_{\rm end}$ premature.
Along the same lines, an identified source with ${\rm SNR}< \eta_{\rm end}$ is not discarded if it is an isolated case that is not in 
a chain of such SNRs. 
 The second criterion 
leads to a substantial saving of computational cost provided it is reached first. Fortunately, this happens to be the case for LDC1-4 where 
the first criterion was used in only $21$
out of $1491$ bands.

% The definition of the inner product in Eq.~\ref{eq:wgninnprod}, hence the SNR in Eq.~\ref{eq:snr},
% uses the instrumental noise PSD for normalization. To shift to the confusion noise or to
% account for uncertainty in the knowledge of the instrumental PSD, a fudge factor $\gamma > 1$
% may be introduced to amplify the PSD. However, $\gamma$ does not affect the 
% sources identified in \fitalgoname: it does not affect the estimation of the
% signal amplitude, $\overline{a}$, because it will cancel out in Eq.~\ref{eq:amp_est}; if $\eta_{\rm end}$ is rescaled to $\eta_{\rm end}/\sqrt{\gamma}$, the sources found before termination will remain the same.

%%%%%%%%%%%%%%%%%%%%%%%%%%%%%

\subsection{Cross-validation}
\label{sec:cv}
A distinguishing feature of \fitalgoname is the 
mitigation of spurious sources by the cross-validation of identified ones. This is implemented by conducting two independent
runs, called primary and secondary, of the iterative source estimation and subtraction in which the search ranges for $\dot{f}$ are set differently. The range used, denoted as $[\dot{f}_{\rm min},\dot{f}_{\rm max}]$, in the primary run is set such that
$\dot{f}_{\rm min} <\dot{f}_{\rm min}^{(0)}$, $\dot{f}_{\rm max} > \dot{f}_{\rm max}^{(0)}$, 
where $\dot{f}_{\rm min}^{(0)}$ and $\dot{f}_{\rm max}^{(0)}>f_{\rm min}^{(0)}$ bound the astrophysically plausible range of $\dot{f}$. Once
the range is set for the primary run, the only requirement on the 
$\dot{f}$ range in the secondary run is that it should be substantially different.

Next, for
the set of estimated parameters $\Theta_a = \{\widehat{\theta}_{a,j}\}$, $j = 1,\ldots,M_a$, of identified sources 
from the primary ($a=1$) and secondary ($a=2$) runs, we
compute, 
\begin{align}
    R_{\rm ee}(\widehat{\theta}_{1,i}) & = \max_j R(\widehat{\theta}_{1,i},\widehat{\theta}_{2,j})\;.
\end{align}
A high value of $R_{\rm ee}$ indicates that an identified source is
a genuine source since it appears in both of the independent searches with similar signal waveforms,
while a low value indicates the contrary.  Thus, 
an identified source from the primary 
run is admitted into the set of reported sources only 
if its $R_{\rm ee}$ value exceeds 
a specified threshold. As shown by our results in Sec.~\ref{sec:srcresolution}, and Fig.~\ref{fig:rdist_ree} in particular, the use of cross-validation as defined above is highly effective in weeding out 
spurious sources. 
%%%%%%%%%%%%%%%%%%%%%%%%%%%%%

%%%%%%%%%%%%%%%%%%%%%%%%%%%%%%%%%%%%%%%%%%%%%%%%%%%%%%%%%%%%%%
\section{Results}
\label{sec:results}
In this section we describe 
the settings for \fitalgoname, the metrics used for assessing its performance, the results
obtained, and the computational costs
associated with the current code.

%%%%%%%%%%%%%%%%%%%%%%%%%%%%%%%
\subsection{Settings}
\label{sec:settings}
The user-specified settings for \fitalgoname and their values for the runs on
LDC1-4 and MLDC-3.1mod are listed below.
\begin{itemize}
    \item{Width of search bands:} While a  bandwidth $B_i$ that 
adapts to the local density of sources is preferable for the 
GB resolution problem, the current version of \fitalgoname keeps $B_i = B$, 
a constant, for the sake of a simpler code.
We set $B = 0.02$~mHz and the starting 
frequency of the first band at $0.09$~mHz. The former is primarily based on considerations of parallelization on the computing clusters that were used, while the latter is the minimum source frequency in LDC1-4.
	%%%%%%
    \item{Termination settings:} As per the description of the termination
    criterion in Sec.~\ref{sec:stoprules}, we set $N_{\rm term}=200$ and  $\eta_{\rm end}\simeq 7$~\cite{Note2} for all the bands. 
These settings were determined empirically, along with the number of search bands above,  to keep the computational cost of the whole search within reasonable bounds.
    %%%%%%
    \item{Edge effect mitigation:} The flat part of the Tukey window used for bandlimiting
    the DFT of the data is $0.015$~mHz wide and the acceptance zone is set to $0.01$~mHz. 
    %%%%%%
    \item{Cross-validation:}
    As mentioned in Sec.~\ref{sec:cv}, 
    the search range for $\dot{f}$ in the 
    primary run should be guided by astrophysical expectations. 
    In this paper, we follow the simpler option of using the known $\dot{f}$
    ranges given in Sec.~\ref{sec:data} to 
    set the ranges as $[-10^{-16},10^{-15}]$~${\rm Hz}^2$ and $[-10^{-14},10^{-13}]$~${\rm Hz}^2$ 
    for $f\leq 4$~mHz and $f\in [4, 15]$~mHz, respectively. 
     The secondary run is used in this paper only 
     for sources below $4$~mHz because it reduces computational costs while 
     preserving the impact of cross-validation where it matters the most.
     For the principal results, the secondary run $\dot{f}$ range is set at
     $[-10^{-14},10^{-13}]$~${\rm Hz}^2$. 
     However, we have also explored the efficacy of
     other combinations of primary and secondary search ranges in cross-validation as described later.
     %%%%%%
     \item{SNR and $R_{ee}$ cuts:} The set of reported sources is extracted
     from the set of identified ones based on their estimated SNR  and $R_{ee}$. We consider several combinations of these cuts as 
     discussed in Sec.~\ref{sec:srcresolution}.
\end{itemize}
   The search ranges for $\lambda$ and $\beta$ cover the whole sky:  
$\lambda\in [0, 2\pi]$ and $\beta \in [-\pi/2,\pi/2]$.
The settings for PSO have already been described in Sec.~\ref{sec:pso}.
Elaborating on the $\dot{f}$ ranges for the primary run, we note that  $\simeq 99.99\%$ of LDC1-4 
sources with $f<4$~mHz have $\dot{f}$ values
in $\simeq [-7.2\times 10^{-18},6.2\times 10^{-17}]$~${\rm Hz}^2$, 
while $\gtrsim 99.9\%$ of them 
have $\dot{f}$ values in  $\simeq [-3.0\times 10^{-15},2.4\times 10^{-14}]$~${\rm Hz}^2$ for $f\in [4,15]$~mHz.  The corresponding numbers for MLDC-3.1 are substantially similar.
Thus, the $\dot{f}$
ranges for the primary run are actually much wider than the
respective ranges for the vast majority of true sources.
%%%%%%%%%%%%%%%%%%%%%%%%%%%%%%

\subsection{Performance metrics}
\label{sec:metrics}
As indicated earlier, the standard metric~\cite{Babak:2007zd} used to judge the performance of a GB 
resolution method is the detection rate: this is the fraction of reported sources
that pass a test of association with true sources. The test used in this 
paper follows the one in MLDC-3~\cite{babak2010mock} and is defined below.

For each reported source, $\widehat{\theta}$, 
the true source $\theta$ is found which (a) has an ${\rm SNR} \geq 3$, (b) has a frequency within $6$ DFT frequencies of $\widehat{f}$, and (c) has the lowest distance, 
defined as $\sum_I(\|\overline{s}(\widehat{\theta})-\overline{s}(\theta)\|^I)^2$, from the
reported source. A reported source is 
confirmed if $R(\theta,\widehat{\theta}) \geq 0.9$.

There are variations of the above test in the literature. In~\cite{Littenberg:2011zg}, the cutoff value for the ${\rm SNR}$ is $\gtrsim 1$ and no mention is made
of a requirement on frequency difference. In~\cite{Blaut:2009si}, the SNR criteria  is implemented by restricting
the true sources to the brightest $\approx 40,000$, 
the frequency separation is reduced to $1$ DFT frequency,
and the minimization of distance 
is replaced by the maximization of $R(\theta,\widehat{\theta})$.  

It can happen occasionally that the same true source becomes the best match,
in terms of the $R(\theta,\widehat{\theta})$ value, for multiple
reported sources.   This wrinkle in assessing GB resolution  
was identified in~\cite{Blaut:2009si}
and handled by confirming
only the source with the highest $R$ (which may be $< 0.9$) from such an ambiguous set. We apply the same
remedy but only for reported sources in the ambiguous set 
that have $R(\theta,\widehat{\theta})\geq 0.9$ for the same true source -- if this threshold is not crossed, none of the reported sources in 
the ambiguous set are confirmed. While
this reduces the number of confirmed sources, the count of reported sources is not touched, thereby resulting in a conservative estimate of the detection rate. 
In practice, we found this to be a negligible issue:  
for example, in one set of $\approx 12,000$ reported sources only a single true source, 
matching a pair of reported sources with $R(\theta,\widehat{\theta})>0.9$, was found.

For assessing the parameter estimation performance of \fitalgoname, we follow the convention common
to past challenges and consider only the mismatch of 
parameters between pairs of confirmed and true sources. 

%%%%%%%%%%%%%%%%%%%%%%%%%%%%%%

\subsection{Source resolution performance}
\label{sec:srcresolution}
\begin{figure}
    \centering
    \includegraphics[width=\columnwidth]{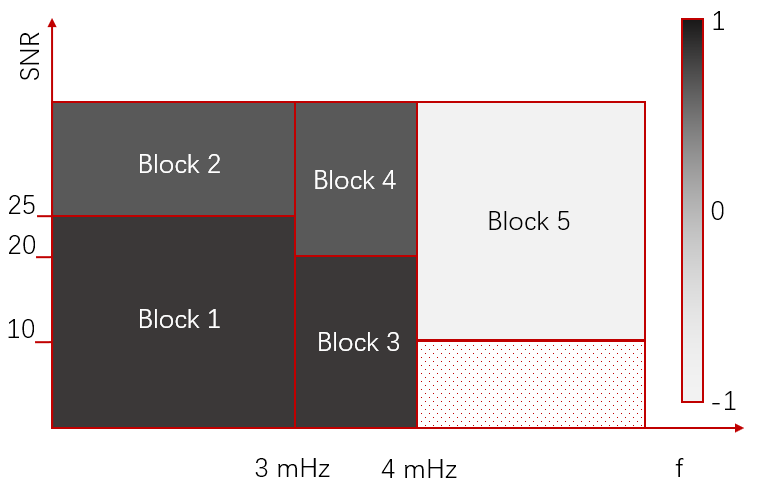}
    \caption{Schematic diagram of the cuts used for extracting reported sources from 
    identified ones. The SNR-frequency plane is partitioned into blocks
    and an identified source falling into a given block is subjected to an $R_{\rm ee}$ 
    threshold indicated by the shading of the block: the correspondence between
    the shading and the numerical value of $R_{\rm ee}$ is shown in the bar at right. 
    The boundaries defining the blocks in this figure are for illustration only. (Note
     an $R_{\rm ee}$ cut is actually not used in this paper for $f\geq 4$~mHz.)}
    \label{fig:snr_ree_freq_block}
\end{figure}

The extraction of reported sources from the set of identified ones is governed 
by SNR and $R_{\rm ee}$ cuts  (c.f., Sec.~\ref{sec:cv}). 
We have the freedom to impose different SNR cuts
in different frequency intervals, as well as different $R_{\rm ee}$
cuts for identified sources above and below a given SNR threshold.
The latter is guided by the expectation that the fraction of spurious sources should increase
as one goes lower in SNR and, hence, the
$R_{\rm ee}$ cutoff to weed them out must be set higher.   
Fig.~\ref{fig:snr_ree_freq_block}
provides a schematic diagram of how the SNR and $R_{\rm ee}$ cuts are implemented: the SNR-frequency plane is partitioned into rectangular {\em blocks} and each block has an associated $R_{\rm ee}$ cutoff that is applied to the identified sources in that block. 

Each combination of SNR and $R_{\rm ee}$ cuts in a given frequency 
interval provides a different  
trade off between the detection 
rate and the search depth -- the minimum SNR found for confirmed sources -- for that interval. 
Thus, by slicing and dicing the same set of identified sources with different combinations of cuts,
we can generate
sets of reported sources that are tailored for specific science goals.
In some cases, such as electromagnetic follow ups, one may want this set to have as low a 
contamination from spurious sources as possible in order to reduce wasted telescope time. In others, such as the estimation of the GB population
distribution, completeness of the reported sample
in terms of SNR may be an important consideration while the 
tolerance to spurious sources could be higher. 

To determine the suitability of a given combination of frequency intervals
and their corresponding SNR and $R_{\rm ee}$ cuts,  robust estimates of the
expected search depths and detection rates
as well as their associated uncertainties would be required. 
These can be obtained from an ensemble of  
mock data realizations drawn from a range of 
realistic astrophysical models. 
While such an exhaustive analysis is outside the scope of this paper, it motivates
a preliminary exploration of the effect of different combinations of cuts
using the two mock data realizations we have in hand. 

  Table~\ref{tab:snr_ree_results(LDC1_TEST1)} presents the results for several different combinations of SNR and $R_{\rm ee}$ cuts for 
  LDC1-4. The corresponding results for MLDC-3.1mod are given in Table~\ref{tab:snr_ree_results_(MLDC3_TEST1)}. The combinations of cuts
  are labeled as follows and they are identical across LDC1-4 and MLDC-3.1mod.
  \begin{itemize}
      \item $R_{\rm ee}$--OFF: No $R_{\rm ee}$ cut used and the only cut applied is on SNR. Hence, cross-validation is not used at all 
      in filtering out reported sources from the set of identified ones. 
      \item $R_{\rm ee}$--ON: The same SNR cuts as in $R_{\rm ee}$--OFF but  $R_{\rm ee}$ cuts are initiated in two out of the 4 blocks below $4$~mHz.
      \item SNR--UP: Same as  $R_{\rm ee}$--ON but $R_{\rm ee}$ cuts are applied to all the blocks below $4$~mHz. 
      \item SNR--DOWN: Same $R_{\rm ee}$ cuts as in SNR--UP but the SNR cut in each block is lower.
      \item MAIN: A combination of higher SNR but lower $R_{\rm ee}$ cuts relative to SNR--UP 
      that provides our principal reported results.  
  \end{itemize}
  In these tables, since $R_{\rm ee}\in [-1,1]$ by definition,
  $R_{\rm ee}=-1$ for a block indicates that no $R_{\rm ee}$ cutoff was applied
  while $R_{\rm ee}=1$ implies that all identified sources were rejected.
The choice of frequency intervals, while being adjustable in principle,
is kept the same in this paper for all of the above combinations.

Comparing $R_{\rm ee}$--OFF and $R_{\rm ee}$--ON in Table~\ref{tab:snr_ree_results(LDC1_TEST1)}, we see how cross-validation improves
the detection rate. The improvement is modest because the $R_{\rm ee}$ cut is used only in blocks $2$ and $4$,
where the identified source SNRs are already high  and, hence, the contamination from 
spurious sources is low. The importance of cross-validation becomes apparent
when the $R_{\rm ee}$ cut is extended to blocks $1$ and $3$
in SNR--UP: we get  $1,468$ 
and $981$ additional reported and confirmed sources, respectively. Lowering the SNR cuts in each block
yields more confirmed sources in SNR--DOWN albeit with a slightly reduced detection rate.

Finally, for MAIN, comprised of higher SNR cuts (relative to SNR--UP) and lower $R_{\rm ee}$ cuts, \fitalgoname recovers  $10,341$ confirmed sources in LDC1-4 -- the largest
among all the combinations -- with an overall
detection rate of $84.28\%$. 
While the search depth for the $[3,4]$~mHz band in MAIN is the same as that for SNR--UP/DOWN, it
improves significantly, from $9.0$ to $7.5$, at frequencies below $3$~mHz.
As seen for the $R_{\rm ee}$--ON combination, it is possible to reach an overall detection rate of $\gtrsim 90\%$ if 
identified 
sources in blocks 1 and 3 are discarded. While this reduces the total number of 
confirmed sources, 
$\approx 8,500$ were still recovered
across the $[0,15]$~mHz band. 

The results for MLDC-3.1mod can be analyzed in a similar fashion as
above. Here, we simply highlight the fact that both the total number of confirmed sources, $10,136$,
as well as the overall detection rate, $84.16\%$, for MAIN in Table~\ref{tab:snr_ree_results_(MLDC3_TEST1)} are 
similar to those for LDC1-4. This is not surprising even though MLDC-3.1mod
has a much larger number of GBs because, as shown in Fig.~\ref{fig:A-f}, most
of the excess sources have low amplitudes and, hence, are not resolvable.

\begin{figure}
    \centering
   \includegraphics[width=\columnwidth]{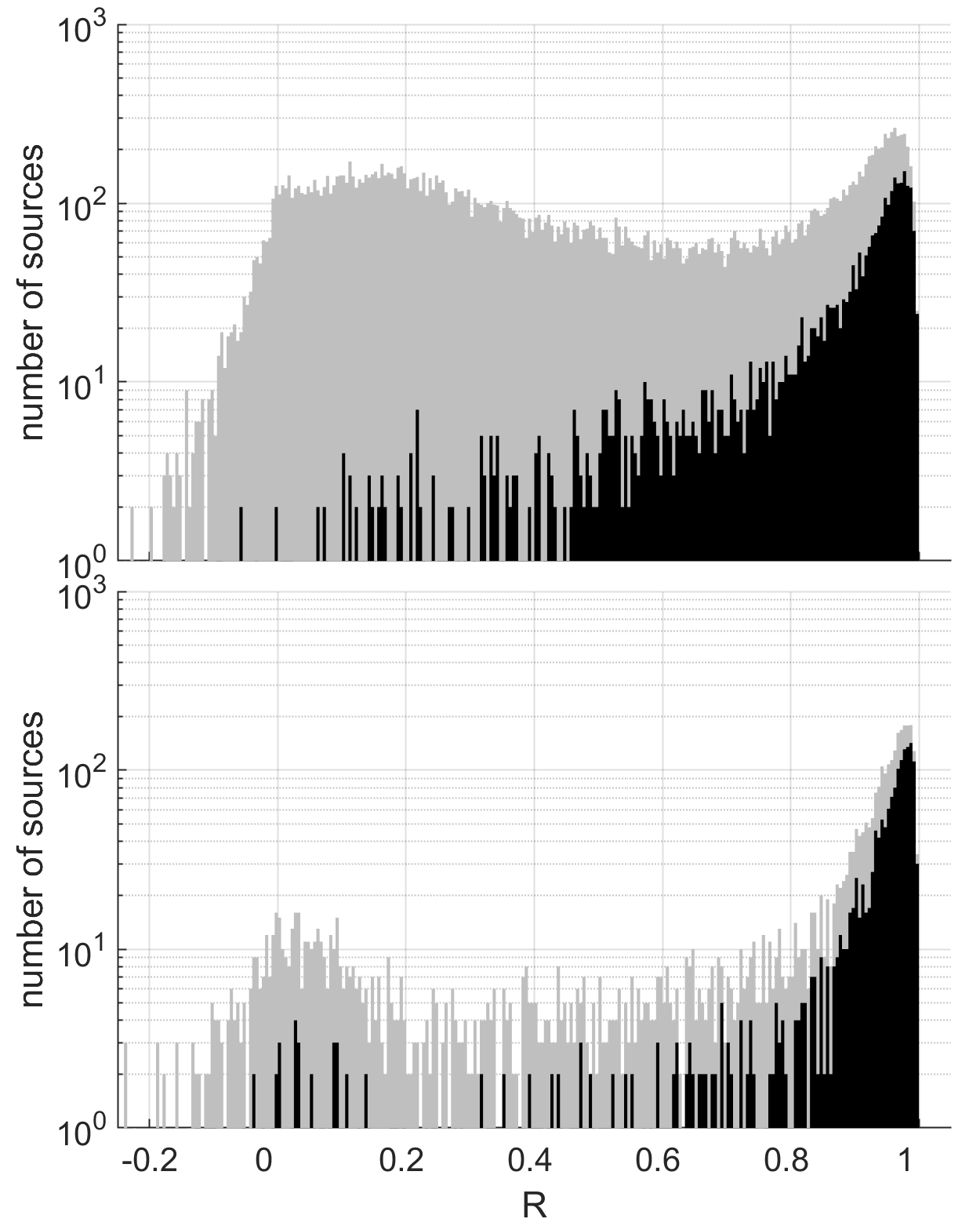}
    \caption{
    Histograms of the correlation coefficient between reported
    and true sources (c.f., Sec.~\ref{sec:metrics}) with and without cross-validation. 
    The top and bottom panels correspond to blocks $1$ and $3$, respectively, for the 
    combination of cuts called MAIN in Table~\ref{tab:snr_ree_results(LDC1_TEST1)}.
    The darker shaded histograms  correspond to the reported 
    sources as obtained with the cuts for these blocks while the lighter
    shaded ones correspond to reported sources for these blocks with the $R_{\rm ee}$ cut removed.
    Without the  $R_{\rm ee}$ cut, the detection rates for blocks 1 and 3 fall 
    from their values given in Table~\ref{tab:snr_ree_results(LDC1_TEST1)} to $17.51\%$
    and $57.26\%$, respectively.
    The $R_{\rm ee}$ cut is seen to predominantly affect the
    reported sources having low correlations that are also more
    likely to be spurious.}
    \label{fig:rdist_ree}
\end{figure}

The effectiveness 
of cross-validation in improving the quality of reported sources is  
demonstrated in Fig.~\ref{fig:rdist_ree}. Not having 
an $R_{\rm ee}$ cut 
leads to a large excess of reported sources that have low 
correlations with true sources: following the discussion 
in Sec.~\ref{sec:metrics}, these are more likely to be 
spurious sources. The effectiveness of the $R_{\rm ee}$ cut
is most striking for the frequency interval below $3$~mHz, where
source confusion is at its highest. While there is some loss
of confirmed sources due to the $R_{\rm ee}$ cut, the 
much larger reduction in spurious sources 
boosts the detection rate significantly.

The above remarks relate to the 
SNR and $R_{\rm ee}$ cuts within the framework of a 
cross-validation scheme that used a much wider 
secondary $\dot{f}$ search range than the primary one. 
Table.~\ref{tab:snr_ree_results_auxcv} contains results from 
an exploration of alternative schemes. In one,
the secondary $\dot{f}$ range is 
reduced by a factor of $10$ with no change to the primary. In the other, the secondary range is brought down to a fixed value,
$\dot{f}=0$, while the primary range is expanded by a factor of $100$. 
The  SNR and $R_{\rm ee}$ cuts are the same in both the cases and  identical to those for the SNR--UP combination in Table.~\ref{tab:snr_ree_results(LDC1_TEST1)}. 

We see that 
reducing
the secondary range by a factor of $10$ increases the total number of 
confirmed sources to $\approx 11,000$ but  worsens the detection rate. This
is most notable in block~1, where it falls from $59.69\%$ 
(c.f., SNR--UP in Table.~\ref{tab:snr_ree_results(LDC1_TEST1)}) to $48.05\%$. 
Remarkably, removing $\dot{f}$ as a search parameter altogether in 
the secondary run does not invalidate the effectiveness of cross-validation. In fact, it significantly improves the detection rate in block~1 but incurs a reduced number of confirmed sources
and worsening of the search depth. 

Our results indicate that it may be possible in a future version of
\fitalgoname 
to refine the cross-validation scheme further by using a mix of secondary runs, with 
different search ranges for $\dot{f}$ 
adapted to different frequency ranges. 

A final remark here is on the role of the assumed noise PSD in 
source resolution performance. In this paper, the noise PSD in the
definition of SNR is that of the instrumental noise. This means that an identified source with SNR near the termination threshold (Sec.~\ref{sec:stoprules}) is not strong relative to the dominant noise, namely confusion, at low frequencies. A uniform SNR termination threshold for all
frequency bands leads, therefore, to more spurious identified sources at lower frequencies. However, the final set of reported sources is determined by the SNR and $R_{\rm ee}$ cuts, which are independent of the termination threshold, and these 
can be adjusted to account for higher noise where needed. (There is a degeneracy between the value used for an SNR cut and the choice of the PSD used in its definition -- either one or both can be tuned for extracting reported sources.) 
%Similarly, blocks 1 and 3
% could be defined with explicit lower bounds on the SNR, further decoupling the reported sources from the termination threshold and the PSD assumed for it.
% In particular, 
% setting an explicit lower bound on SNR for blocks $1$ and $3$ would effectively decouple
% their detection rates from the noise PSD used in the termination threshold. 
% (For illustration, setting a lower bound of ${\rm SNR} = 15$ in block $1$ would improve the detection rate to
% $68.12\%$ for MAIN in Table~\ref{tab:snr_ree_results(LDC1_TEST1)} without any change in the termination threshold.) 
% In this sense, the influence of the assumed noise PSD on the set of reported sources and the detection rate is quite limited. 

%\input{tables_new}
%%%%%%%%% FIRST TABLE %%%%%%%%%%%%%%%%%
\begin{table*}
    \centering
    \begin{tabular}{|c|c|c|c|c|c|c|c|c|c|c|}
    %%%%%%%%%%%%%%R_{\rm ee}$:OFF%%%%%%%%%%%%%%%%%%%%
    \hline
        \multirow{3}{*}{$R_{\rm ee}$--OFF} & \multicolumn{2}{|c|}{Block 1} & \multicolumn{2}{|c|}{Block 2} & 
        \multicolumn{2}{|c|}{Block 3} & \multicolumn{2}{|c|}{Block 4} & \multicolumn{2}{|c|}{Block 5}\\
    \cline{2-11}
        &$\nu$~mHz &  SNR & $\nu$~mHz &  SNR & $\nu$~mHz &  SNR & $\nu$~mHz &  SNR &  $\nu$~mHz &  SNR  \\
    \cline{2-11}
        &$[0,3]$ & $[0,20]$ & $[0,3]$ & $[20,\infty]$ & $[3,4]$ & $[0,15]$ 
        &$[3,4]$ & $[15,\infty]$ & $[4,15]$ & $[10,\infty]$ \\
    \hline
   $R_{\rm ee}$ & \multicolumn{2}{|c|}{$1$} & \multicolumn{2}{|c|}{$-1$} &
               \multicolumn{2}{|c|}{$1$} & \multicolumn{2}{|c|}{$-1$} & \multicolumn{2}{|c|}{$-1$}\\
   \hline
   Identified   & \multicolumn{2}{|c|}{$20397$} & \multicolumn{2}{|c|}{$3254$} &
               \multicolumn{2}{|c|}{$2741$} & \multicolumn{2}{|c|}{$2314$} & \multicolumn{2}{|c|}{$4281$}\\
   \hline
   Reported     & \multicolumn{2}{|c|}{$0$} & \multicolumn{2}{|c|}{$3254$} &
               \multicolumn{2}{|c|}{$0$} & \multicolumn{2}{|c|}{$2314$} & \multicolumn{2}{|c|}{$4281$}\\
   \hline
   Detection rate  & \multicolumn{2}{|c|}{$  0\%$} & \multicolumn{2}{|c|}{$83.53\%$} &
               \multicolumn{2}{|c|}{$  0\%$} & \multicolumn{2}{|c|}{$87.51\%$} & \multicolumn{2}{|c|}{$94.23\%$}\\
   \hline
   Lowest SNR (confirmed)  & \multicolumn{4}{|c|}{$20.0$} & \multicolumn{4}{|c|}{$15.0$} &
               \multicolumn{2}{|c|}{$10.0$}\\
    \hline
    \multicolumn{1}{|c|}{Total reported} & \multicolumn{10}{|c|}{$9849$}\\
    \hline
    \multicolumn{1}{|c|}{Total confirmed} & \multicolumn{10}{|c|}{$8777$}\\
    \hline
    \multicolumn{1}{|c|}{Detection rate} & \multicolumn{10}{|c|}{$89.12\%$}\\
    
    %%%%%%%%%%%%%%$R_{\rm ee}$:ON%%%%%%%%%%%%%%%%%%%%
    \hline
        \multirow{3}{*}{$R_{\rm ee}$--ON} & \multicolumn{2}{|c|}{Block 1} & \multicolumn{2}{|c|}{Block 2} & 
        \multicolumn{2}{|c|}{Block 3} & \multicolumn{2}{|c|}{Block 4} & \multicolumn{2}{|c|}{Block 5}\\
    \cline{2-11}
        &$\nu$~mHz &  SNR & $\nu$~mHz &  SNR & $\nu$~mHz &  SNR & $\nu$~mHz &  SNR &  $\nu$~mHz &  SNR  \\
    \cline{2-11}
        &$[0,3]$ & $[0,20]$ & $[0,3]$ & $[20,\infty]$ & $[3,4]$ & $[0,15]$ 
        &$[3,4]$ & $[15,\infty]$ & $[4,15]$ & $[10,\infty]$ \\
    \hline
   $R_{\rm ee}$ & \multicolumn{2}{|c|}{$1$} & \multicolumn{2}{|c|}{$0.9$} &
               \multicolumn{2}{|c|}{$1$} & \multicolumn{2}{|c|}{$0.9$} & \multicolumn{2}{|c|}{$-1$}\\
   \hline
   Identified   & \multicolumn{2}{|c|}{$20397$} & \multicolumn{2}{|c|}{$3254$} &
               \multicolumn{2}{|c|}{$2741$} & \multicolumn{2}{|c|}{$2314$} & \multicolumn{2}{|c|}{$4281$}\\
   \hline
   Reported     & \multicolumn{2}{|c|}{$0$} & \multicolumn{2}{|c|}{$2886$} &
               \multicolumn{2}{|c|}{$0$} & \multicolumn{2}{|c|}{$2124$} & \multicolumn{2}{|c|}{$4281$}\\
   \hline
   Detection rate  & \multicolumn{2}{|c|}{$  0\%$} & \multicolumn{2}{|c|}{$87.21\%$} &
               \multicolumn{2}{|c|}{$  0\%$} & \multicolumn{2}{|c|}{$90.44\%$} & \multicolumn{2}{|c|}{$94.23\%$}\\
   \hline
   Lowest SNR (confirmed)  & \multicolumn{4}{|c|}{$20.0$} &
                             \multicolumn{4}{|c|}{$15.0$} & 
                             \multicolumn{2}{|c|}{$10.0$}\\
    \hline
    \multicolumn{1}{|c|}{Total reported} & \multicolumn{10}{|c|}{$9291$}\\
    \hline
    \multicolumn{1}{|c|}{Total confirmed} & \multicolumn{10}{|c|}{$8472$}\\
    \hline
    \multicolumn{1}{|c|}{Detection rate} & \multicolumn{10}{|c|}{$91.19\%$}\\
    
  %%%%%%%%%%%%%%SNR:UP%%%%%%%%%%%%%%%%%%%%
    \hline
    \multirow{3}{*}{SNR--UP} &\multicolumn{2}{|c|}{Block 1}& \multicolumn{2}{|c|}{Block 2}& \multicolumn{2}{|c|}{Block 3}&
    \multicolumn{2}{|c|}{Block 4}& \multicolumn{2}{|c|}{Block 5}\\
    \cline{2-11}
        &$\nu$~mHz &  SNR & $\nu$~mHz &  SNR & $\nu$~mHz &  SNR & $\nu$~mHz &  SNR &  $\nu$~mHz &  SNR  \\
    \cline{2-11}
        &$[0,3]$ & $[0,20]$ & $[0,3]$ & $[20,\infty]$ & $[3,4]$ & $[0,15]$ 
        &$[3,4]$ & $[15,\infty]$ & $[4,15]$ & $[10,\infty]$ \\
    \hline
   $R_{\rm ee}$ & \multicolumn{2}{|c|}{$0.99$} & \multicolumn{2}{|c|}{$0.9$} &
               \multicolumn{2}{|c|}{$0.99$} & \multicolumn{2}{|c|}{$0.9$} & \multicolumn{2}{|c|}{$-1$}\\
   \hline
   Identified   & \multicolumn{2}{|c|}{$20397$} & \multicolumn{2}{|c|}{$3254$} &
               \multicolumn{2}{|c|}{$2741$} & \multicolumn{2}{|c|}{$2314$} & \multicolumn{2}{|c|}{$4281$}\\
   \hline
   Reported     & \multicolumn{2}{|c|}{$918$} & \multicolumn{2}{|c|}{$2886$} &
               \multicolumn{2}{|c|}{$550$} & \multicolumn{2}{|c|}{$2124$} & \multicolumn{2}{|c|}{$4281$}\\
   \hline
   Detection rate  & \multicolumn{2}{|c|}{$59.69\%$} & \multicolumn{2}{|c|}{$87.21\%$} &
               \multicolumn{2}{|c|}{$78.73\%$} & \multicolumn{2}{|c|}{$90.44\%$} & \multicolumn{2}{|c|}{$94.23\%$}\\
   \hline
   Lowest SNR (confirmed)  & \multicolumn{4}{|c|}{$9.0$} &
                             \multicolumn{4}{|c|}{$6.6$} &
                             \multicolumn{2}{|c|}{$10.0$}\\
    \hline
    \multicolumn{1}{|c|}{Total reported} & \multicolumn{10}{|c|}{$10759$}\\
    \hline
    \multicolumn{1}{|c|}{Total confirmed} & \multicolumn{10}{|c|}{$9453$}\\
    \hline
    \multicolumn{1}{|c|}{Detection rate} & \multicolumn{10}{|c|}{$87.86\%$}\\
    
    %%%%%%%%%%%%%%SNR:DOWN%%%%%%%%%%%%%%%%%%%%
    \hline
        \multirow{3}{*}{SNR--DOWN} & \multicolumn{2}{|c|}{Block 1} & \multicolumn{2}{|c|}{Block 2} & 
        \multicolumn{2}{|c|}{Block 3} & \multicolumn{2}{|c|}{Block 4} & \multicolumn{2}{|c|}{Block 5}\\
    \cline{2-11}
        &$\nu$~mHz &  SNR & $\nu$~mHz &  SNR & $\nu$~mHz &  SNR & $\nu$~mHz &  SNR &  $\nu$~mHz &  SNR  \\
    \cline{2-11}
        &$[0,3]$ & $[0,15]$ & $[0,3]$ & $[15,\infty]$ & $[3,4]$ & $[0,10]$ 
        &$[3,4]$ & $[10,\infty]$ & $[4,15]$ & $[8,\infty]$ \\
    \hline
   $R_{\rm ee}$ & \multicolumn{2}{|c|}{$0.99$} & \multicolumn{2}{|c|}{$0.9$} &
               \multicolumn{2}{|c|}{$0.99$} & \multicolumn{2}{|c|}{$0.9$} & \multicolumn{2}{|c|}{$-1$}\\
   \hline
   Identified   & \multicolumn{2}{|c|}{$18001$} & \multicolumn{2}{|c|}{$5650$} &
               \multicolumn{2}{|c|}{$1462$} & \multicolumn{2}{|c|}{$3593$} & \multicolumn{2}{|c|}{$4550$}\\
   \hline
   Reported     & \multicolumn{2}{|c|}{$258$} & \multicolumn{2}{|c|}{$4016$} &
               \multicolumn{2}{|c|}{$99$} & \multicolumn{2}{|c|}{$2874$} & \multicolumn{2}{|c|}{$4550$}\\
   \hline
   Detection rate  & \multicolumn{2}{|c|}{$51.16\%$} & \multicolumn{2}{|c|}{$80.13\%$} &
               \multicolumn{2}{|c|}{$63.64\%$} & \multicolumn{2}{|c|}{$87.72\%$} & \multicolumn{2}{|c|}{$90.55\%$}\\
   \hline
   Lowest SNR (confirmed)  & \multicolumn{4}{|c|}{$9.0$} &
                             \multicolumn{4}{|c|}{$6.6$} &
                             \multicolumn{2}{|c|}{$8.0$}\\
    \hline
    \multicolumn{1}{|c|}{Total reported} & \multicolumn{10}{|c|}{$11797$}\\
    \hline
    \multicolumn{1}{|c|}{Total confirmed} & \multicolumn{10}{|c|}{$10054$}\\
    \hline
    \multicolumn{1}{|c|}{Detection rate} & \multicolumn{10}{|c|}{$85.23\%$}\\
  %%%%%%%%%%%%%%MAIN%%%%%%%%%%%%%%%%%%%%
    \hline
        \multirow{3}{*}{MAIN} & \multicolumn{2}{|c|}{Block 1} & \multicolumn{2}{|c|}{Block 2} & 
        \multicolumn{2}{|c|}{Block 3} & \multicolumn{2}{|c|}{Block 4} & \multicolumn{2}{|c|}{Block 5}\\
    \cline{2-11}
        &$\nu$~mHz &  SNR & $\nu$~mHz &  SNR & $\nu$~mHz &  SNR & $\nu$~mHz &  SNR &  $\nu$~mHz &  SNR  \\
    \cline{2-11}
        &$[0,3]$ & $[0,25]$ & $[0,3]$ & $[25,\infty]$ & $[3,4]$ & $[0,20]$ 
        &$[3,4]$ & $[20,\infty]$ & $[4,15]$ & $[10,\infty]$ \\
    \hline
   $R_{\rm ee}$ & \multicolumn{2}{|c|}{$0.9$} & \multicolumn{2}{|c|}{$0.5$} &
               \multicolumn{2}{|c|}{$0.9$} & \multicolumn{2}{|c|}{$0.5$} & \multicolumn{2}{|c|}{$-1$}\\
   \hline
   Identified   & \multicolumn{2}{|c|}{$21546$} & \multicolumn{2}{|c|}{$2105$} &
               \multicolumn{2}{|c|}{$3531$} & \multicolumn{2}{|c|}{$1524$} & \multicolumn{2}{|c|}{$4281$}\\
   \hline
   Reported     & \multicolumn{2}{|c|}{$2778$} & \multicolumn{2}{|c|}{$2072$} &
               \multicolumn{2}{|c|}{$1629$} & \multicolumn{2}{|c|}{$1510$} & \multicolumn{2}{|c|}{$4281$}\\
   \hline
   Detection rate  & \multicolumn{2}{|c|}{$62.56\%$} & \multicolumn{2}{|c|}{$91.17\%$} &
               \multicolumn{2}{|c|}{$79.13\%$} & \multicolumn{2}{|c|}{$92.12\%$} & \multicolumn{2}{|c|}{$94.23\%$}\\
   \hline
   Lowest SNR (confirmed)  & \multicolumn{4}{|c|}{$7.5$}  &
                              \multicolumn{4}{|c|}{$6.6$} & 
                             \multicolumn{2}{|c|}{$10.0$}\\
    \hline
    \multicolumn{1}{|c|}{Total reported} & \multicolumn{10}{|c|}{$12270$}\\
    \hline
    \multicolumn{1}{|c|}{Total confirmed} & \multicolumn{10}{|c|}{$10341$}\\
    \hline
    \multicolumn{1}{|c|}{Detection rate} & \multicolumn{10}{|c|}{$84.28\%$}\\
    
    \hline
    \end{tabular}
    \caption{
    Performance of \fitalgoname for different combinations of 
    SNR and $R_{\rm ee}$ cuts on LDC1-4 data. For cross-validation (implemented for $f\leq 4$~mHz),
    the primary and 
    secondary search ranges for $\dot{f}$ are 
    $[-10^{-16},  10^{-15}]~\rm{Hz^2}$ and $[-10^{-14}, 10^{-13}]~\rm{Hz^2}$ respectively.  A block for which 
    the $R_{\rm ee}$ cut was not used is shown as having $R_{\rm ee}=-1$, while $R_{\rm ee}=1$ means that the identified sources in that block were discarded.
    }
    \label{tab:snr_ree_results(LDC1_TEST1)}
\end{table*}

%%%%%%%%% SECOND TABLE %%%%%%%%%%%%%%%%%
\begin{table*}
    \centering
    \begin{tabular}{|c|c|c|c|c|c|c|c|c|c|c|}

    %%%%%%%%%%%%%%$R_{\rm ee}$:OFF%%%%%%%%%%%%%%%%%%%%
    \hline
        \multirow{3}{*}{$R_{\rm ee}$--OFF} & \multicolumn{2}{|c|}{Block 1} & \multicolumn{2}{|c|}{Block 2} & 
        \multicolumn{2}{|c|}{Block 3} & \multicolumn{2}{|c|}{Block 4} & \multicolumn{2}{|c|}{Block 5}\\
    \cline{2-11}
        &$\nu$~mHz &  SNR & $\nu$~mHz &  SNR & $\nu$~mHz &  SNR & $\nu$~mHz &  SNR &  $\nu$~mHz &  SNR  \\
    \cline{2-11}
        &$[0,3]$ & $[0,20]$ & $[0,3]$ & $[20,\infty]$ & $[3,4]$ & $[0,15]$ 
        &$[3,4]$ & $[15,\infty]$ & $[4,15]$ & $[10,\infty]$ \\
    \hline
   $R_{\rm ee}$ & \multicolumn{2}{|c|}{$1$} & \multicolumn{2}{|c|}{$-1$} &
               \multicolumn{2}{|c|}{$1$} & \multicolumn{2}{|c|}{$-1$} & \multicolumn{2}{|c|}{$-1$}\\
   \hline
   Identified   & \multicolumn{2}{|c|}{$20914$} & \multicolumn{2}{|c|}{$3253$} &
               \multicolumn{2}{|c|}{$3412$} & \multicolumn{2}{|c|}{$2204$} & \multicolumn{2}{|c|}{$4582$}\\
   \hline
   Reported     & \multicolumn{2}{|c|}{$0$} & \multicolumn{2}{|c|}{$3253$} &
               \multicolumn{2}{|c|}{$0$} & \multicolumn{2}{|c|}{$2204$} & \multicolumn{2}{|c|}{$4582$}\\
   \hline
   Detection rate  & \multicolumn{2}{|c|}{$  0\%$} & \multicolumn{2}{|c|}{$84.69\%$} &
               \multicolumn{2}{|c|}{$  0\%$} & \multicolumn{2}{|c|}{$86.21\%$} & \multicolumn{2}{|c|}{$91.58\%$}\\
   \hline
   Lowest SNR (confirmed)  & \multicolumn{4}{|c|}{$20.0$} &
                              \multicolumn{4}{|c|}{$15.0$} &
                              \multicolumn{2}{|c|}{$10.0$}\\
    \hline
    \multicolumn{1}{|c|}{Total reported} & \multicolumn{10}{|c|}{$10039$}\\
    \hline
    \multicolumn{1}{|c|}{Total confirmed} & \multicolumn{10}{|c|}{$8851$}\\
    \hline
    \multicolumn{1}{|c|}{Detection rate} & \multicolumn{10}{|c|}{$88.17\%$}\\

    %%%%%%%%%%%%%%$R_{\rm ee}$:ON%%%%%%%%%%%%%%%%%%%%
    \hline
    	\multirow{3}{*}{$R_{\rm ee}$--ON} & \multicolumn{2}{|c|}{Block 1} & \multicolumn{2}{|c|}{Block 2} & 
        \multicolumn{2}{|c|}{Block 3} & \multicolumn{2}{|c|}{Block 4} & \multicolumn{2}{|c|}{Block 5}\\
    \cline{2-11}
        &$\nu$~mHz &  SNR & $\nu$~mHz &  SNR & $\nu$~mHz &  SNR & $\nu$~mHz &  SNR &  $\nu$~mHz &  SNR  \\
    \cline{2-11}
        &$[0,3]$ & $[0,20]$ & $[0,3]$ & $[20,\infty]$ & $[3,4]$ & $[0,15]$ 
        &$[3,4]$ & $[15,\infty]$ & $[4,15]$ & $[10,\infty]$ \\
    \hline
   $R_{\rm ee}$ & \multicolumn{2}{|c|}{$1$} & \multicolumn{2}{|c|}{$0.9$} &
               \multicolumn{2}{|c|}{$1$} & \multicolumn{2}{|c|}{$0.9$} & \multicolumn{2}{|c|}{$-1$}\\
   \hline
   Identified   & \multicolumn{2}{|c|}{$20914$} & \multicolumn{2}{|c|}{$3253$} &
               \multicolumn{2}{|c|}{$3412$} & \multicolumn{2}{|c|}{$2204$} & \multicolumn{2}{|c|}{$4582$}\\
   \hline
   Reported     & \multicolumn{2}{|c|}{$0$} & \multicolumn{2}{|c|}{$2868$} &
               \multicolumn{2}{|c|}{$0$} & \multicolumn{2}{|c|}{$1937$} & \multicolumn{2}{|c|}{$4582$}\\
   \hline
   Detection rate  & \multicolumn{2}{|c|}{$  0\%$} & \multicolumn{2}{|c|}{$88.18\%$} &
               \multicolumn{2}{|c|}{$  0\%$} & \multicolumn{2}{|c|}{$90.24\%$} & \multicolumn{2}{|c|}{$91.58\%$}\\
   \hline
   Lowest SNR (confirmed)  & \multicolumn{4}{|c|}{$20.0$} &
                             \multicolumn{4}{|c|}{$15.0$} &
                             \multicolumn{2}{|c|}{$10.0$}\\
    \hline
    \multicolumn{1}{|c|}{Total reported} & \multicolumn{10}{|c|}{$9387$}\\
    \hline
    \multicolumn{1}{|c|}{Total confirmed} & \multicolumn{10}{|c|}{$8473$}\\
    \hline
    \multicolumn{1}{|c|}{Detection rate} & \multicolumn{10}{|c|}{$90.26\%$}\\

    %%%%%%%%%%%%%%SNR:UP%%%%%%%%%%%%%%%%%%%%
    \hline
        \multirow{3}{*}{SNR--UP} & \multicolumn{2}{|c|}{Block 1} & \multicolumn{2}{|c|}{Block 2} & 
        \multicolumn{2}{|c|}{Block 3} & \multicolumn{2}{|c|}{Block 4} & \multicolumn{2}{|c|}{Block 5}\\
    \cline{2-11}
        &$\nu$~mHz &  SNR & $\nu$~mHz &  SNR & $\nu$~mHz &  SNR & $\nu$~mHz &  SNR &  $\nu$~mHz &  SNR  \\
    \cline{2-11}
        &$[0,3]$ & $[0,20]$ & $[0,3]$ & $[20,\infty]$ & $[3,4]$ & $[0,15]$ 
        &$[3,4]$ & $[15,\infty]$ & $[4,15]$ & $[10,\infty]$ \\
    \hline
   $R_{\rm ee}$ & \multicolumn{2}{|c|}{$0.99$} & \multicolumn{2}{|c|}{$0.9$} &
               \multicolumn{2}{|c|}{$0.99$} & \multicolumn{2}{|c|}{$0.9$} & \multicolumn{2}{|c|}{$-1$}\\
   \hline
   Identified   & \multicolumn{2}{|c|}{$20914$} & \multicolumn{2}{|c|}{$3253$} &
               \multicolumn{2}{|c|}{$3412$} & \multicolumn{2}{|c|}{$2204$} & \multicolumn{2}{|c|}{$4582$}\\
   \hline
   Reported     & \multicolumn{2}{|c|}{$797$} & \multicolumn{2}{|c|}{$2868$} &
               \multicolumn{2}{|c|}{$427$} & \multicolumn{2}{|c|}{$1937$} & \multicolumn{2}{|c|}{$4582$}\\
   \hline
   Detection rate  & \multicolumn{2}{|c|}{$62.99\%$} & \multicolumn{2}{|c|}{$88.18\%$} &
               \multicolumn{2}{|c|}{$74.00\%$} & \multicolumn{2}{|c|}{$90.24\%$} & \multicolumn{2}{|c|}{$91.58\%$}\\
   \hline
   Lowest SNR (confirmed)  & \multicolumn{4}{|c|}{$7.8$} &
                             \multicolumn{4}{|c|}{$7.7$} &
                            \multicolumn{2}{|c|}{$10.0$}\\
    \hline
    \multicolumn{1}{|c|}{Total reported} & \multicolumn{10}{|c|}{$10611$}\\
    \hline
    \multicolumn{1}{|c|}{Total confirmed} & \multicolumn{10}{|c|}{$9291$}\\
    \hline
    \multicolumn{1}{|c|}{Detection rate} & \multicolumn{10}{|c|}{$87.56\%$}\\
    
    %%%%%%%%%%%%%%SNR:DOWN%%%%%%%%%%%%%%%%%%%%
    \hline
        \multirow{3}{*}{SNR--DOWN} & \multicolumn{2}{|c|}{Block 1} & \multicolumn{2}{|c|}{Block 2} & 
        \multicolumn{2}{|c|}{Block 3} & \multicolumn{2}{|c|}{Block 4} & \multicolumn{2}{|c|}{Block 5}\\
    \cline{2-11}
        &$\nu$~mHz &  SNR & $\nu$~mHz &  SNR & $\nu$~mHz &  SNR & $\nu$~mHz &  SNR &  $\nu$~mHz &  SNR  \\
    \cline{2-11}
        &$[0,3]$ & $[0,15]$ & $[0,3]$ & $[15,\infty]$ & $[3,4]$ & $[0,10]$ 
        &$[3,4]$ & $[10,\infty]$ & $[4,15]$ & $[8,\infty]$ \\
    \hline
   $R_{\rm ee}$ & \multicolumn{2}{|c|}{$0.99$} & \multicolumn{2}{|c|}{$0.9$} &
               \multicolumn{2}{|c|}{$0.99$} & \multicolumn{2}{|c|}{$0.9$} & \multicolumn{2}{|c|}{$-1$}\\
   \hline
   Identified   & \multicolumn{2}{|c|}{$18499$} & \multicolumn{2}{|c|}{$5668$} &
               \multicolumn{2}{|c|}{$1874$} & \multicolumn{2}{|c|}{$3742$} & \multicolumn{2}{|c|}{$5055$}\\
   \hline
   Reported     & \multicolumn{2}{|c|}{$202$} & \multicolumn{2}{|c|}{$3977$} &
               \multicolumn{2}{|c|}{$57$} & \multicolumn{2}{|c|}{$2631$} & \multicolumn{2}{|c|}{$5055$}\\
   \hline
   Detection rate  & \multicolumn{2}{|c|}{$51.98\%$} & \multicolumn{2}{|c|}{$80.94\%$} &
               \multicolumn{2}{|c|}{$61.40\%$} & \multicolumn{2}{|c|}{$86.66\%$} & \multicolumn{2}{|c|}{$85.68\%$}\\
   \hline
   Lowest SNR (confirmed)  & \multicolumn{4}{|c|}{$7.8$} &
                             \multicolumn{4}{|c|}{$7.7$} & 
                             \multicolumn{2}{|c|}{$8.0$}\\
    \hline
    \multicolumn{1}{|c|}{Total reported} & \multicolumn{10}{|c|}{$11922$}\\
    \hline
    \multicolumn{1}{|c|}{Total confirmed} & \multicolumn{10}{|c|}{$9970$}\\
    \hline
    \multicolumn{1}{|c|}{Detection rate} & \multicolumn{10}{|c|}{$83.63\%$}\\
 
     %%%%%%%%%%%%%%MAIN%%%%%%%%%%%%%%%%%%%%
    \hline
        \multirow{3}{*}{MAIN} & \multicolumn{2}{|c|}{Block 1} & \multicolumn{2}{|c|}{Block 2} & 
        \multicolumn{2}{|c|}{Block 3} & \multicolumn{2}{|c|}{Block 4} & \multicolumn{2}{|c|}{Block 5}\\
    \cline{2-11}
        &$\nu$~mHz &  SNR & $\nu$~mHz &  SNR & $\nu$~mHz &  SNR & $\nu$~mHz &  SNR &  $\nu$~mHz &  SNR  \\
    \cline{2-11}
        &$[0,3]$ & $[0,25]$ & $[0,3]$ & $[25,\infty]$ & $[3,4]$ & $[0,20]$ 
        &$[3,4]$ & $[20,\infty]$ & $[4,15]$ & $[10,\infty]$ \\
    \hline
   $R_{\rm ee}$ & \multicolumn{2}{|c|}{$0.9$} & \multicolumn{2}{|c|}{$0.5$} &
               \multicolumn{2}{|c|}{$0.9$} & \multicolumn{2}{|c|}{$0.5$} & \multicolumn{2}{|c|}{$-1$}\\
   \hline
   Identified   & \multicolumn{2}{|c|}{$22054$} & \multicolumn{2}{|c|}{$2113$} &
               \multicolumn{2}{|c|}{$4223$} & \multicolumn{2}{|c|}{$1393$} & \multicolumn{2}{|c|}{$4582$}\\
   \hline
   Reported     & \multicolumn{2}{|c|}{$2553$} & \multicolumn{2}{|c|}{$2096$} &
               \multicolumn{2}{|c|}{$1443$} & \multicolumn{2}{|c|}{$1370$} & \multicolumn{2}{|c|}{$4582$}\\
   \hline
   Detection rate  & \multicolumn{2}{|c|}{$62.48\%$} & \multicolumn{2}{|c|}{$92.56\%$} &
               \multicolumn{2}{|c|}{$78.79\%$} & \multicolumn{2}{|c|}{$92.55\%$} & \multicolumn{2}{|c|}{$91.58\%$}\\
   \hline
   Lowest SNR (confirmed)  & \multicolumn{4}{|c|}{$7.8$} &
                           \multicolumn{4}{|c|}{$7.7$}   & 
                           \multicolumn{2}{|c|}{$10.0$}\\
    \hline
    \multicolumn{1}{|c|}{Total reported} & \multicolumn{10}{|c|}{$12044$}\\
    \hline
    \multicolumn{1}{|c|}{Total confirmed} & \multicolumn{10}{|c|}{$10136$}\\
    \hline
    \multicolumn{1}{|c|}{Detection rate} & \multicolumn{10}{|c|}{$84.16\%$}\\
   %%%%%%%%%%%%%%%%%%%%%%%%%%%%%%%%%%%%%%%%   
    \hline
    \end{tabular}
    \caption{
    Performance of \fitalgoname for different combinations of 
    SNR and $R_{\rm ee}$ cuts on MLDC-3.1mod data. For cross-validation (implemented for $f\leq 4$~mHz),
    the primary and 
    secondary search ranges for $\dot{f}$ are 
    $[-10^{-16},  10^{-15}]~\rm{Hz^2}$ and $[-10^{-14}, 10^{-13}]~\rm{Hz^2}$ respectively. A block for which 
    the $R_{\rm ee}$ cut was not used is shown as having $R_{\rm ee}=-1$, while $R_{\rm ee}=1$ means that the identified sources in that block were discarded.
    %MLDC3\underline{~~}\{TEST1\underline{~~}MLDC3\}: The two $\dot{f}$ search ranges below $4$~mHz are $[-10^{-16}, 10^{-15}]~\rm{Hz^2}$ and $[-10^{-14}, 10^{-13}]~\rm{Hz^2}$ respectively. The former one is used as ``base", while the latter is as ``comparison".
    }
    \label{tab:snr_ree_results_(MLDC3_TEST1)}
\end{table*}

%%%%%%%%% THIRD TABLE %%%%%%%%%%%%%%%%%
\begin{table*}
    \centering
    \begin{tabular}{|c|c|c|c|c|c|c|c|c|c|c|}
    \hline
        \multirow{3}{*}{$\begin{array}{lr}{\rm primary:}&[-10^{-16}, 10^{-15}]~\rm{Hz^2}\\{\rm secondary:}&[-10^{-15}, 10^{-14}]~\rm{Hz^2}\end{array}$} 
        & \multicolumn{2}{|c|}{Block 1} & \multicolumn{2}{|c|}{Block 2} & 
        \multicolumn{2}{|c|}{Block 3} & \multicolumn{2}{|c|}{Block 4} & \multicolumn{2}{|c|}{Block 5}\\
    \cline{2-11}
    %%%%%%%%%%%%%%Set 1%%%%%%%%%%%%%%%%%%%%
    %\hline
        &$\nu$~mHz &  SNR & $\nu$~mHz &  SNR & $\nu$~mHz &  SNR & $\nu$~mHz &  SNR &  $\nu$~mHz &  SNR  \\
    \cline{2-11}
        &$[0,3]$ & $[0,20]$ & $[0,3]$ & $[20,\infty]$ & $[3,4]$ & $[0,15]$ 
        &$[3,4]$ & $[15,\infty]$ & $[4,15]$ & $[10,\infty]$ \\
    \hline
   $R_{\rm ee}$ & \multicolumn{2}{|c|}{$0.99$} & \multicolumn{2}{|c|}{$0.9$} &
               \multicolumn{2}{|c|}{$0.99$} & \multicolumn{2}{|c|}{$0.9$} & \multicolumn{2}{|c|}{$-1$}\\
   \hline
   Identified   & \multicolumn{2}{|c|}{$20397$} & \multicolumn{2}{|c|}{$3254$} &
               \multicolumn{2}{|c|}{$2741$} & \multicolumn{2}{|c|}{$2314$} & \multicolumn{2}{|c|}{$4281$}\\
   \hline
   Reported     & \multicolumn{2}{|c|}{$3003$} & \multicolumn{2}{|c|}{$3114$} &
               \multicolumn{2}{|c|}{$1198$} & \multicolumn{2}{|c|}{$2235$} & \multicolumn{2}{|c|}{$4281$}\\
   \hline
   Detection rate  & \multicolumn{2}{|c|}{$48.05\%$} & \multicolumn{2}{|c|}{$85.36\%$} &
               \multicolumn{2}{|c|}{$71.87\%$} & \multicolumn{2}{|c|}{$89.08\%$} & \multicolumn{2}{|c|}{$94.23\%$}\\
   \hline
   Lowest SNR (confirmed)  & \multicolumn{4}{|c|}{$7.2$} &
                             \multicolumn{4}{|c|}{$7.0$} &
                             \multicolumn{2}{|c|}{$10.0$}\\
    \hline
    \multicolumn{1}{|c|}{Total reported} & \multicolumn{10}{|c|}{$13831$}\\
    \hline
    \multicolumn{1}{|c|}{Total confirmed} & \multicolumn{10}{|c|}{$10987$}\\
    \hline
    \multicolumn{1}{|c|}{Detection rate} & \multicolumn{10}{|c|}{$79.44\%$}\\
    %%%%%%%%%%%%%%%%%%%%%%%%%%%%%%%%%%%%%%%%%%%%%%%%%%%%%%%%%%%%%%%%%%%%%%
    \hline
    % &\multicolumn{2}{|c|}{Block 1}& \multicolumn{2}{|c|}{Block 2}& \multicolumn{2}{|c|}{Block 3}&
    % \multicolumn{2}{|c|}{Block 4}& \multicolumn{2}{|c|}{Block 5}\\
        \multirow{3}{*}{$\begin{array}{lr}{\rm primary:}&[-10^{-14}, 10^{-13}]~\rm{Hz^2}\\{\rm secondary:}&[0, 0]~\rm{Hz^2}\end{array}$} 
        & \multicolumn{2}{|c|}{Block 1} & \multicolumn{2}{|c|}{Block 2} & 
        \multicolumn{2}{|c|}{Block 3} & \multicolumn{2}{|c|}{Block 4} & \multicolumn{2}{|c|}{Block 5}\\
    \cline{2-11}
    %%%%%%%%%%%%%%Set 1%%%%%%%%%%%%%%%%%%%%
    % \hline
        &$\nu$~mHz &  SNR & $\nu$~mHz &  SNR & $\nu$~mHz &  SNR & $\nu$~mHz &  SNR &  $\nu$~mHz &  SNR  \\
    \cline{2-11}
        &$[0,3]$ & $[0,20]$ & $[0,3]$ & $[20,\infty]$ & $[3,4]$ & $[0,15]$ 
        &$[3,4]$ & $[15,\infty]$ & $[4,15]$ & $[10,\infty]$ \\
    \hline
   $R_{\rm ee}$ & \multicolumn{2}{|c|}{$0.99$} & \multicolumn{2}{|c|}{$0.9$} &
               \multicolumn{2}{|c|}{$0.99$} & \multicolumn{2}{|c|}{$0.9$} & \multicolumn{2}{|c|}{$-1$}\\
   \hline
   Identified   & \multicolumn{2}{|c|}{$22149$} & \multicolumn{2}{|c|}{$3147$} &
               \multicolumn{2}{|c|}{$3021$} & \multicolumn{2}{|c|}{$2284$} & \multicolumn{2}{|c|}{$4281$}\\
   \hline
   Reported     & \multicolumn{2}{|c|}{$570$} & \multicolumn{2}{|c|}{$2752$} &
               \multicolumn{2}{|c|}{$328$} & \multicolumn{2}{|c|}{$2037$} & \multicolumn{2}{|c|}{$4281$}\\
   \hline
   Detection rate  & \multicolumn{2}{|c|}{$71.58\%$} & \multicolumn{2}{|c|}{$87.72\%$} &
               \multicolumn{2}{|c|}{$86.28\%$} & \multicolumn{2}{|c|}{$91.65\%$} & \multicolumn{2}{|c|}{$94.23\%$}\\
   \hline
   Lowest SNR (confirmed)  & \multicolumn{4}{|c|}{$9.0$} &
                             \multicolumn{4}{|c|}{$6.9$} & 
                             \multicolumn{2}{|c|}{$10.0$}\\
    \hline
    \multicolumn{1}{|c|}{Total reported} & \multicolumn{10}{|c|}{$9968$}\\
    \hline
    \multicolumn{1}{|c|}{Total confirmed} & \multicolumn{10}{|c|}{$9006$}\\
    \hline
    \multicolumn{1}{|c|}{Detection rate} & \multicolumn{10}{|c|}{$90.35\%$}\\
    \hline
    \end{tabular}
    \caption{
    Performance of \fitalgoname for the same combination of 
    SNR and $R_{\rm ee}$ cuts on LDC1-4 data but
    different choices for $\dot{f}$ search ranges in the primary and secondary runs as listed in 
    the heading for each combination. A block for which 
    the $R_{\rm ee}$ cut was not used is shown as having $R_{\rm ee}=-1$.
    }
    \label{tab:snr_ree_results_auxcv}
\end{table*}
%%%%%%%%%%%%%%%%%%%%%%%%%%%%%%

%%%%%%%%%%%%%%%%%%%%%%%%%%%%%%
\subsection{Residuals}
\label{sec:residual}
Resolving GBs is not only important in and of itself but also crucial to the extraction of other types of sources from LISA data. 
This calls for an examination of the residual data in a TDI 
combination after subtracting out the 
signals of reported GBs. We do this only for   
 LDC1-4 TDI A  and the reported
sources obtained with the combination of SNR and $R_{\rm ee}$ cuts called MAIN in Table~\ref{tab:snr_ree_results(LDC1_TEST1)}.  The results for TDI E and 
other combinations of cuts are 
visually indistinguishable from the ones shown here.  

\begin{figure*}
    \centering
    \includegraphics[width=\textwidth]{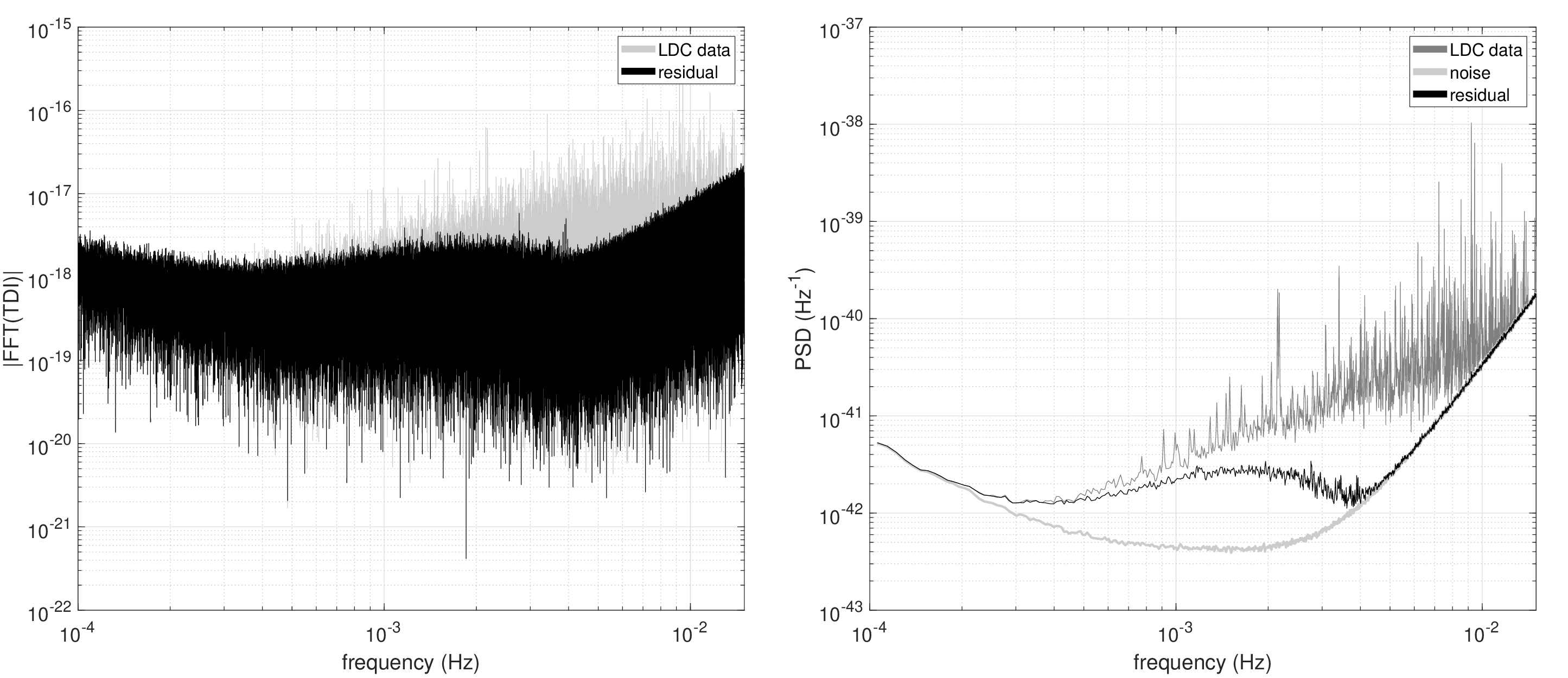}
    \caption{Periodogram (left) and PSD (right) of the residual  after removing the reported sources for the MAIN combination in Table~\ref{tab:snr_ree_results(LDC1_TEST1)} from LDC1-4 TDI A data. 
    The frequency resolution of the PSD is $8.1380\times 10^{-3}$~mHz.}
    \label{fig:Residual_PSD_A}
\end{figure*}

Figure~\ref{fig:Residual_PSD_A} compares
the periodogram (magnitude of the DFT)
and PSD of the residual  with those of the data
and the instrumental noise realization in the data. 
We see that the PSD of the residual 
is  brought down to the level of the instrumental noise for frequencies 
$\gtrsim 4$~mHz. 
In fact, 
 the cross-correlation coefficient between the residual and instrumental noise time series, both bandlimited to $[4,15]$~mHz, is $0.9973$, indicating that they are nearly identical and that practically all resolvable GBs were removed with high accuracy. 

\begin{figure}
    \centering
    \includegraphics[width=\columnwidth]{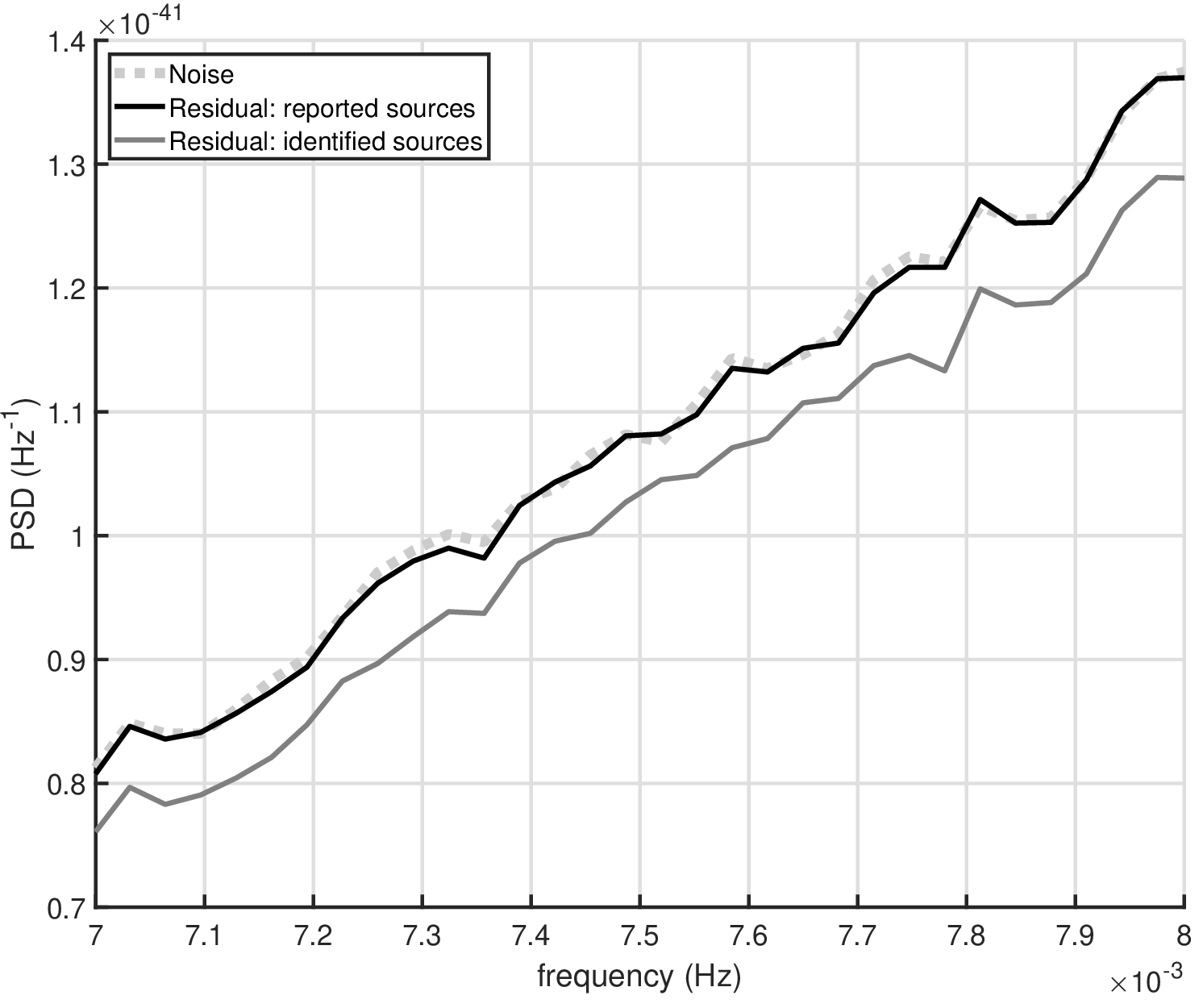}
    \caption{PSD of the residual in the $[4,15]$~mHz range obtained by removing  (black curve) only the reported sources in the MAIN combination of Table~\ref{tab:snr_ree_results(LDC1_TEST1)}, and (gray curve) all the identified sources. Due to a higher fraction of spurious sources in the latter set, the residual PSD  is lowered below that of the 
    instrumental noise (dotted curve). (The displayed frequency
    interval has been restricted to $[7,8]$~mHz for visual clarity.)}
    \label{fig:Residual_PSD_A_overfit}
\end{figure}

It is worth emphasizing here the necessity of comparing the residual with the instrumental noise
and not just the data.
 This is illustrated in Fig.~\ref{fig:Residual_PSD_A_overfit}, where the 
residuals after subtracting the reported and identified 
sources in the $[4,15]$~mHz range are compared. We see that the PSD of the 
residual corresponding to the latter set, which has more spurious sources
by definition, now lies below that of
the instrumental noise. This is clearly an artifact of overfitting that comparing
the PSDs of the residual and data alone would not reveal,
misleading one into thinking that the GBs had been subtracted out
faithfully.

 Starting below $4$~mHz, the residual power increases towards lower 
 frequencies due to the rising confusion between the ever more crowded signals. 
 It is interesting to note in Fig.~\ref{fig:Residual_PSD_A} that the PSD of the
 residual is less spiky compared to that of the data. This is an indication 
 that loud resolvable sources have been taken out successfully.

%%%%%%%%%%%%%%%%%%%%%%%%%%%%%%

\subsection{Parameter estimation performance}
\begin{figure*}
	\centering
	\includegraphics[width=\textwidth]{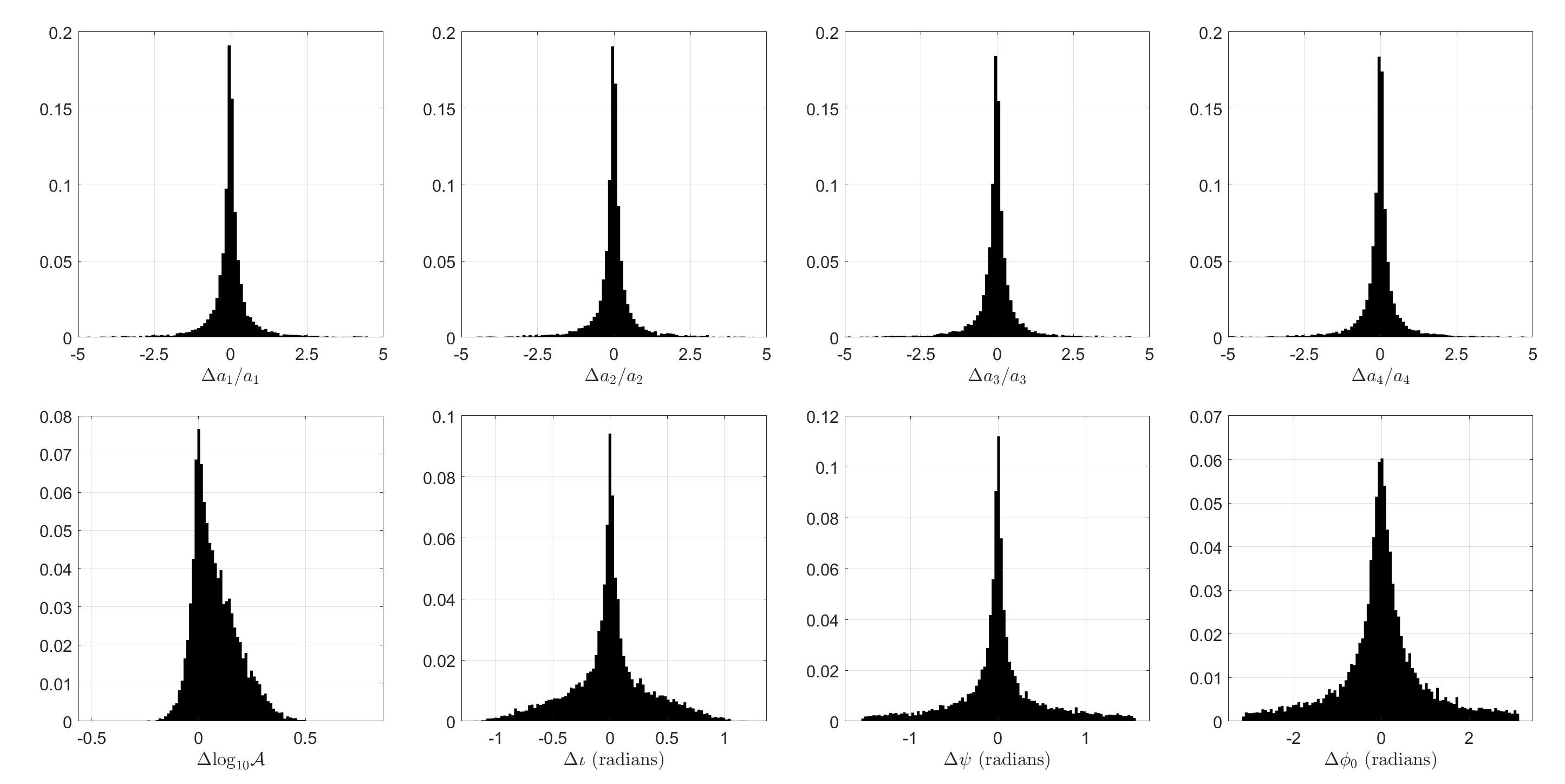}
	\caption{Histogram of the differences between the extrinsic parameters of confirmed and  true sources in LDC1-4 data. The confirmed sources are obtained from the combination of cuts called MAIN in Table~\ref{tab:snr_ree_results(LDC1_TEST1)}. For each distribution, a small 
	fraction of outliers are not shown for visual clarity. The number of outliers dropped, as a fraction of the $10,341$ confirmed sources, for $\Delta a_i$, 
	$i = 1$ to $4$, are $2.6\%$, $2.4\%$, $2.6\%$, and  $2.7\%$, respectively. 
	Note that the extrinsic parameters in the bottom row are functions of the
	ones in the top row.
		\label{fig:extrnsc_errors}}
\end{figure*}

\begin{figure*}
	\centering
	\includegraphics[width=\textwidth]{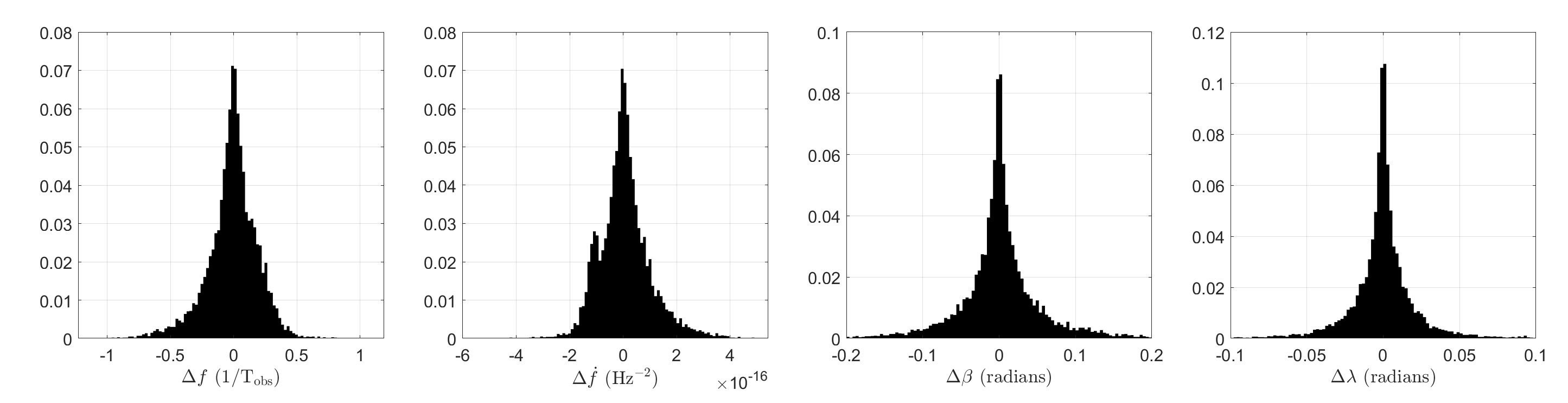}
	\caption{Histogram of the differences between the intrinsic parameters of confirmed and  true sources in LDC1-4 data. The confirmed sources are obtained from the combination of cuts called MAIN in Table~\ref{tab:snr_ree_results(LDC1_TEST1)}. In the distributions of
	$\Delta \beta$ and $\Delta\lambda$, $1.5\%$ and  $1\%$, respectively, of the  $10,341$ confirmed sources have been dropped for visual clarity.}
	\label{fig:intrnsc_errors}
\end{figure*}

Figures~\ref{fig:extrnsc_errors} and~\ref{fig:intrnsc_errors} show the marginal 
distributions of the differences between the parameters of confirmed and true sources. 
They appear to be qualitatively similar to those in~\cite{Blaut:2009si,Littenberg:2011zg}. 

\begin{figure*}
	\centering
	\includegraphics[width=\textwidth]{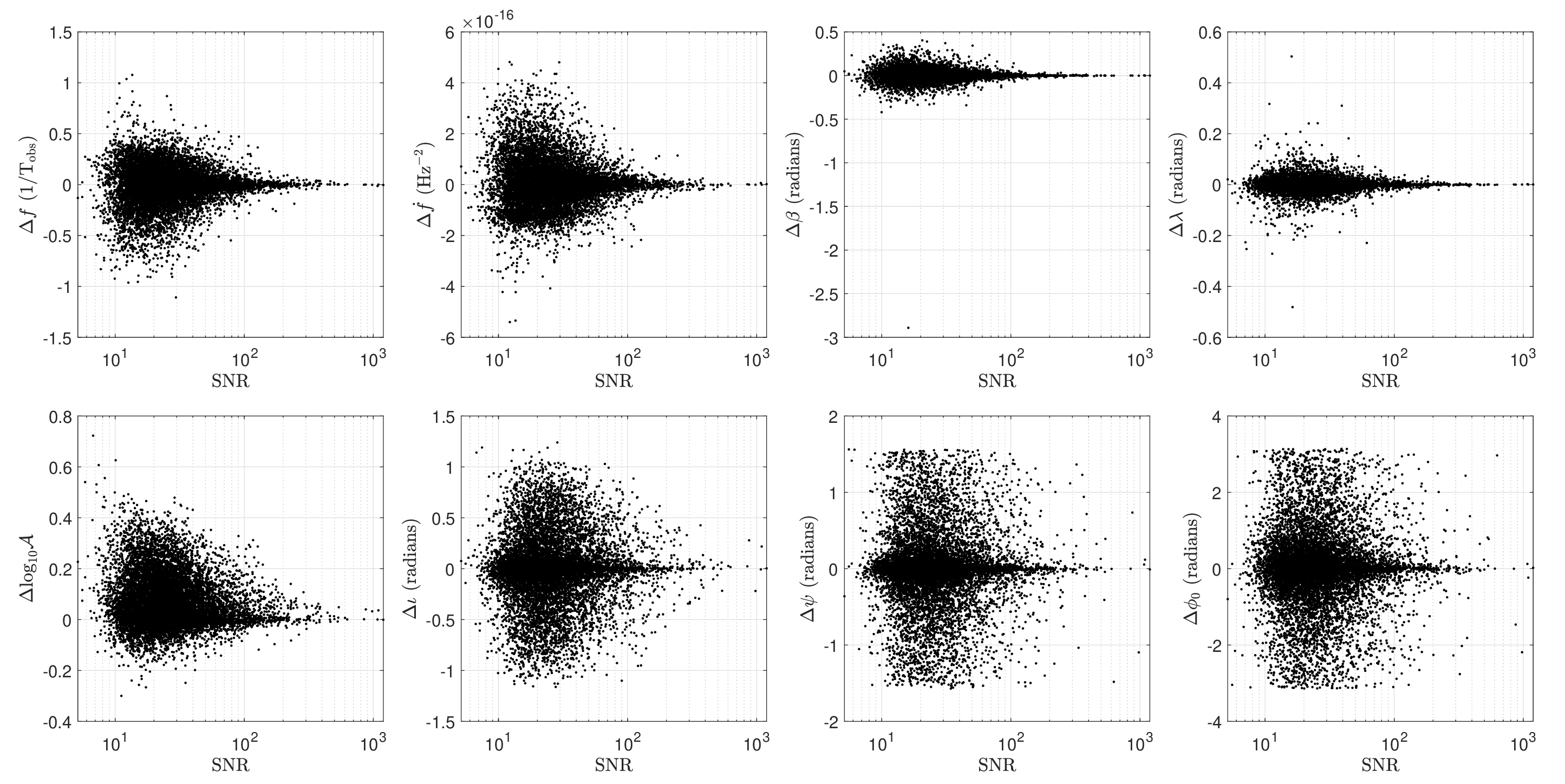}
	\caption{Scatterplots of differences in the parameters of confirmed and true sources in LDC1-4 against the SNR of the latter. The outliers
	 that were excluded in Fig.~\ref{fig:extrnsc_errors} and Fig.~\ref{fig:intrnsc_errors} are included here.}
	\label{fig:scatter_parsSNR}
\end{figure*}

It is worth noting that the differences in parameter values are computed over
 a spread in true source parameters and their distributions are, in general, not the same as those of parameter estimation errors for a fixed true source. (The former are mixture distributions while the latter are not.) In particular, the distribution of parameter differences will be affected by
the distribution of the SNR of resolvable sources. This is illustrated in Fig.~\ref{fig:scatter_parsSNR} containing
scatterplots of the differences  against the true source SNR.
It can be seen that the asymmetry in 
the distribution of $\Delta \log_{10}\mathcal{A}$
is more pronounced for lower SNR sources, which also 
happen to be predominantly at lower frequencies where the confusion between sources is higher.
We conjecture that the higher density of sources 
at lower frequencies  leads
to spurious constructive interference that biases
the estimation of the amplitude of a given source to higher values. 
\label{sec:paramest}

Overall, the estimation of every parameter shows asymptotic convergence to the true parameter
with increasing SNR. This supports the statistical consistency of 
the iterative source subtraction approach to GB resolution.

%%%%%%%%%%%%%%%%%%%%%%%%%%%%%%

\subsection{Computational considerations}
\label{sec:compcost}
Cross-validation requires at least two complete passes of a multi-source
resolution method on the same data. Hence, its computational feasibility hinges
 on the runtime of the  method.
For \fitalgoname, the two main accelerators of runtime 
are the efficient
global optimization of \fstat by PSO and the undersampling of template waveforms. The former reduces the number of  \fstat evaluations while the 
latter speeds up each one.

The current implementation of \fitalgoname is in the Matlab~\cite{MATLAB:2019a}  programming language.  The $\nruns$ independent PSO runs (c.f., Sec.~\ref{sec:pso}) are executed in parallel on a matching number of cores of a multi-core processor, while the frequency bands are analyzed in parallel on different nodes of a computing cluster. For each band and PSO run, the code completes the analysis of $2$~yr of data in 
 $\lesssim 15$~hrs on a $2.3$~GHz core. A third level of parallelization,
 over PSO particle fitness evaluations in each iteration, is possible but this cannot be implemented in Matlab and requires a 
 compiled language such as C. In principle, achieving all three layers of parallelization could speed up the code by up to a factor of $\lesssim 40$, the number of PSO particles. (The shift to a compiled language by itself should
 afford a significant increase in speed.)

%%%%%%%%%%%%%%%%%%%%%%%%%%%%%%%%%%%%%%%%%%%%%%%%%%%%%%%%%%%%%%
\section{Conclusions}
\label{sec:conclusions}
We have presented a new method, \fitalgoname, for resolving Galactic binaries in LISA data and have quantified its performance on LDC1-4 and (modified) MLDC3.1 data. Our results
affirm the validity of the iterative single source estimation and subtraction approach
for the LISA multi-source resolution problem.
The 
principal 
novel features of \fitalgoname are the use of PSO for maximization of
the  \fstat, fast template generation using undersampling, and mitigation of
spurious sources using a cross-validation scheme. We also showed
that the trade-off between the detection rate and the depth of the search
can be tuned using different combinations of the cuts  used for
extracting reported sources from the initial set of identified ones. 

Inspection of the residual left from the subtraction of reported sources shows that it is possible to reach the instrumental noise floor for frequencies $\gtrsim 4$~mHz.  Increased confusion noise below $4$~mHz 
precludes this but the residual power still
remains substantially lower than that of the data down to $\simeq 1$~mHz. 
We envision \fitalgoname to be a part of a hierarchical analysis pipeline where it will be applied to the residual after subtracting louder signals, such as massive black hole binaries, from the data. Provided such signals are reduced below the GB confusion noise, \fitalgoname can take over and find resolvable GBs.  However, quantifying the effects of weaker non-GB sources left behind in the data on a pure GB search pipeline, such as \fitalgoname, remains an open question. The ability to reach the instrumental noise floor above $4$~mHz bodes well for the extraction of non-GB sources in this frequency band although an actual test must await future work.

The performance of \fitalgoname in terms of the 
number of reported sources, the detection rate, and 
 differences between the parameters of confirmed and true sources
 is comparable to state-of-the-art methods. There remains substantial scope for improvements, 
 ranging from minor, such as how 
 PSO handles the search over the sky (currently treated as a box instead of a sphere), to major, 
 such as supplementing iterative single source fitting with a global one that fits multiple sources simultaneously. The latter is also required to address a current limitation, namely, error estimates for the parameters of individual reported sources. Since the errors arise not only from instrumental or confusion noise but also from the presence of other resolvable
 sources, a global fit analysis is required.
 We also need to go beyond our current, empirically obtained, understanding of
  cross-validation and find out how to better control its performance. 

With significantly improved runtimes in the future, it will become possible to
characterize the performance of \fitalgoname 
on an ensemble of realizations of the GB population. This will, in turn, enable a statistically rigorous study of the effect of different combinations of cuts on 
detection rates and search depths. 

%%%%%%%%%%%%%%%%%%%%%%%%%%%%%%%%%%%%%%%%%%%%%%%%%%%%%%%%%%%%%%
\section*{Acknowledgements}
Zhang and Liu are supported by the National Key Research and Development Program of China grant 2020YFC2201400, the 111 Project under grant B20063, and the Fundamental Research Funds for the Central Universities (grants lzujbky-2019-ct06 and  lzujbky-2020-it04).
Group activities are financially supported by the Morningside Center of Mathematics and a part of the MPG-CAS collaboration in
low frequency gravitational wave physics. Partial support from the Xiandao B project in gravitational wave detection is also acknowledged.
We are grateful to Prof. Shing-Tung Yau for his long
term and unconditional support and Prof. Yun-Kau Lau for assembling our research group together.
The contribution of S.D.M. to this paper was partially supported by National Science Foundation (NSF)
 grant PHY-1505861. We thank Shaodong Zhao for help with extracting numbers from published figures.  We gratefully acknowledge the use of high performance computers at (i) the Supercomputing Center of Lanzhou University, (ii) State Key Laboratory of Scientific and Engineering Computing, CAS, and (iii)
 the Texas Advanced Computing Center (TACC) at 
the University of Texas at Austin (www.tacc.utexas.edu). 

%%%%%%%%%%%%%%%%%%%%%%%%%%%%%%%%%%%%%%%%%%%%%%%%%%%%%%%%%%%%%%

\bibliographystyle{ieeetr}
%\bibliography{xiaobo_bib}
%\bibliographystyle{natbib}

\end{document}